\documentclass[5p,twocolumn,times,number]{elsarticle}
\makeatletter
\let\c@author\relax
\makeatother

\expandafter\let\csname ver@natbib.sty\endcsname\relax
\usepackage[backend=biber,style=phys,eprint=true,hyperref=true,biblabel=brackets,block=ragged,natbib=true]{biblatex}
\usepackage{graphicx}
\usepackage{amsmath}   
\usepackage{etoolbox}
\usepackage{amssymb}
\usepackage{epsfig}
\usepackage{lineno, hyperref}
\usepackage{url}
\usepackage{amsthm}
\usepackage[export]{adjustbox}
\usepackage[font=footnotesize,labelfont=bf,skip=7pt]{caption} 
\usepackage{subfig}
\usepackage{subcaption}
\usepackage{xpatch}
\usepackage{xspace}
\usepackage{float}

\hypersetup{
    colorlinks=true,
    linkcolor=blue,
    filecolor=magenta,      
    urlcolor=cyan,
    pdftitle={proANUBIS Calibration},
    pdfpagemode=FullScreen,
    }

\def\pmbanner{}

\newcommand{\proanubis}{\mbox{\textsl{pro}ANUBIS}\xspace}

\newcommand{\madversion}[1]{MG5\_aMC\@NLO #1}

\newcommand{\metre}{\ensuremath{\textnormal{m}}\xspace}

\newcommand{\invfb}{\ensuremath{{\rm fb}^{-1}}\xspace}

\begin{document}
\begin{frontmatter}

\title{ \pmbanner {Calibration and Performance of \proanubis:\\ a proof-of-concept detector for the ANUBIS experiment}\\
\includegraphics[width=0.1\linewidth]{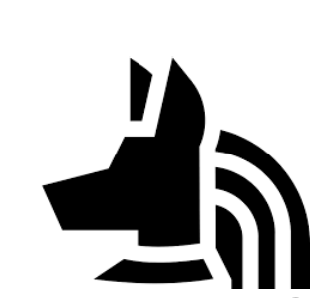}
\\
{\large\textbf{ANUBIS Collaboration}}
\\[-0.2em]
{\normalsize\textit{E-mail: }\href{mailto:anubis-publications@cern.ch}{anubis-publications@cern.ch}}
\\[-0.2em]
}

\author[18]{Giulio Aielli}
\author[1]{Oleg Brandt}
\author[20]{Jude Burling}
\author[20]{Patrick Collins}
\author[1]{Jonas Dej}
\author[1]{Cayetano Fernandez Ruiz}
\author[1]{Christopher Lester}
\author[20]{Kaijia Liu}
\author[9]{Luca Pizzimento}
\author[2]{Ludovico Pontecorvo}
\author[12,20]{Th\'eo Reymermier}
\author[1]{Michael Revering}
\author[1]{Aashaq Shah}
\author[8,20]{Tom Spencer}
\author[1]{Paul Swallow}
\author[1]{Julian Wack}
\author[6,20]{Yanglin Wan}
\author[]{for the ANUBIS Collaboration}
\address[1]{Cavendish Laboratory, University of Cambridge, Cambridge, United Kingdom}
\address[2]{CERN, Geneva, Switzerland}
\address[6]{School of Physics and Astronomy, University of Edinburgh, Edinburgh, United Kingdom}
\address[8]{School of Sciences and Engineering, Glasgow Caledonian University, Glasgow, United Kingdom}
\address[9]{Department of Physics, University of Hong Kong, Hong Kong SAR, China}
\address[12]{Institut de Physique des 2 Infinis (IP2I), Lyon, France}
\address[18]{University and INFN Roma Tor Vergata, Rome, Italy}
\address[20]{Formerly at: Cavendish Laboratory, University of Cambridge, Cambridge, United Kingdom}

\begin{abstract}
Long-lived particles with lifetimes $\tau>10$~ps are predicted by many extensions of the Standard Model with viable dark matter candidates. 
The ANUBIS experiment proposes to extend the experimental sensitivity to long-lived particles by instrumenting the ceiling of the ATLAS cavern with Resistive Plate Chamber detectors in order to reconstruct vertices from long-lived particle decays in the air-filled volume above the ATLAS detector. 
The \proanubis detector has been installed in the ATLAS cavern to validate the detector technology planned for ANUBIS and to take \emph{in-situ} measurements of muon and hadron fluxes inside the ATLAS cavern using $pp$ collision data from the LHC. 
In this paper, the data collected, reconstruction techniques used, and performance of the \proanubis detector are discussed. 
The detection efficiency and timing resolution are found to be consistent with expectations and to meet the performance requirements of ANUBIS.
%
\end{abstract}

\end{frontmatter}
 
\section{Introduction} 
\label{Sec:Intro}
Many extensions of the Standard Model with viable dark matter candidates predict long-lived particles (LLP) with lifetimes $\tau>10$~ps.
The ANUBIS experiment~\cite{Bauer:2019vqk,bib:anubis_v2} proposes to extend the Large Hadron Collider (LHC)~\cite{Evans:1129806} sensitivity to LLPs~\cite{Alimena:2019zri,PBC:2025sny} significantly beyond the reach of existing experiments by instrumenting the ceiling of the UX15 experimental cavern hosting the ATLAS detector~\cite{Aad:1129811,ATLAS:2023dns} with Resistive Plate Chamber (RPC) detectors~\cite{SANTONICO1981377}. 
ANUBIS expects to achieve this improvement in sensitivity by reconstructing decay vertices produced by LLPs in the air-filled volume above ATLAS by measuring the tracks of their charged decay products. This configuration provides both a high signal acceptance due to the large decay volume located above and relatively near the Interaction Point (IP), and low backgrounds due to the low scattering cross section of the air and the ability to use ATLAS as an active veto~\cite{bib:anubis_v2}. 

The performance of the RPC detectors~\cite{Pizzimento_2020,Aad_2021} and their ability to accurately reconstruct displaced vertices from charged particle tracks is a key aspect of the expected ANUBIS sensitivity. 
High detection efficiency is essential for reliably observing signal tracks and forming decay vertices, as well as to differentiate signal vertices from background interactions and detector noise. 
Precise spatial resolution is required for accurate track reconstruction and vertex localisation, thereby enabling tighter vetoes on interactions with material.
Similarly, precise timing resolution can be used to reject downwards-going particles, tracks from slow particles with $\beta\ll 1$, or other non-collision backgrounds, and improve vertex fitting by including time information. To reach the design goals of the ANUBIS detector, the hit time resolution should be $<$~0.5~ns, the spatial resolution should be $<$~5~mm, and the per-tracking-layer efficiency should be $>$~98$\%$~\cite{Bauer:2019vqk}.

To experimentally determine the above performance parameters and thereby validate that the technology can meet the design requirements of ANUBIS, a prototype detector \proanubis was constructed~\cite{ANUBIS:2025mme} in 2022, installed in the ATLAS cavern and commissioned in 2023--24~\cite{ANUBIS:2025inn}. 
The positioning and design of \proanubis is also expected to allow the measurement of Standard Model LLPs interacting in the air volume between ATLAS and ANUBIS, hence providing an empirical benchmark for simulation studies of background processes.

\section {The \proanubis detector}
\label{Sec:proanubis}
The \proanubis detector is designed to replicate a single module of the planned ANUBIS experiment, and consists of three tracking layers with per-layer horizontal dimensions of 0.99~\metre $\times$ 1.78~\metre, where the layers are separated vertically with a spacing of 0.6~\metre between the lowest pair of tracking layers and 0.5~\metre between the upper pair. 
Each tracking layer can contain up to three individual BIS7S~\cite{Massa:2020hjw} RPC detectors, or ‘planes'. 
The bottom tracking layer is instrumented with three RPC detectors, the middle layer contains a single RPC, and the top layer contains two RPC detectors. 
This configuration, rather than two triplets separated by 1~\metre, was chosen to assess the impact of a more-centrally positioned tracking layer on the track and vertex reconstruction performance, which will be studied in future publications. 
The detector used the same gas mixture as the RPCs in the ATLAS muon system, which in 2024 was  64$\%$
C$_2$H$_2$F$_4$, 30$\%$ CO$_2$, 5$\%$ i-C$_4$H$_{10}$, 1$\%$ SF$_6$~\cite{SIMSEK2025170618} and in 2025 was 64.5$\%$
C$_2$H$_2$F$_4$, 30$\%$ CO$_2$, 5$\%$ i-C$_4$H$_{10}$, 0.5$\%$ SF$_6$.

Referring to the number of RPCs contained in each layer, the bottom, middle, and top layers are referred to as the Triplet, Singlet, and Doublet, respectively. 
Within the triplet, the individual RPCs are labelled as ‘Low’, ‘Mid’, and ‘Top’, corresponding to their vertical positions where applicable. 
Similarly, the RPCs in the doublet are labelled as ‘Low' and ‘Top'.

The RPC detector itself consists of a 1~mm gas gap equipped with readout panels with 64 copper signal pickup strips running parallel to the short edge of the detector and 32 strips parallel to the long side of the detector on the other side. 
A full description of the construction and initial performance of \proanubis can be found in Ref.~\cite{ANUBIS:2025mme}, and its installation within the ATLAS cavern and data readout is described in~\cite{ANUBIS:2025inn}.

\begin{figure}[ht]
\centering
\includegraphics[width=0.45\textwidth]{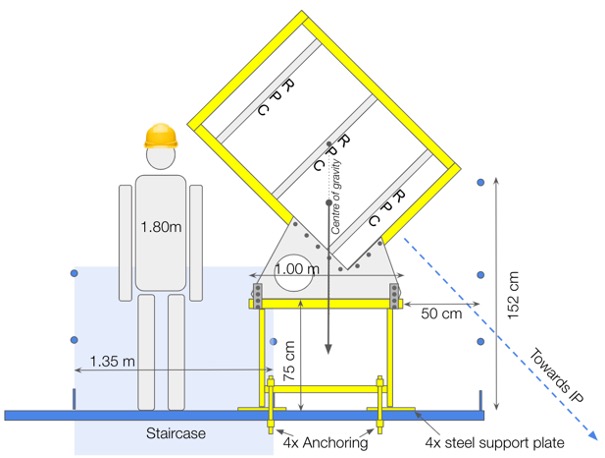}
\caption{A sketch of the \proanubis detector showing the three RPC tracking layers along with the support frame.} 
\label{fig:proAnubLayout}
\end{figure}
\begin{figure}[ht]
\centering
\includegraphics[width=0.45\textwidth]{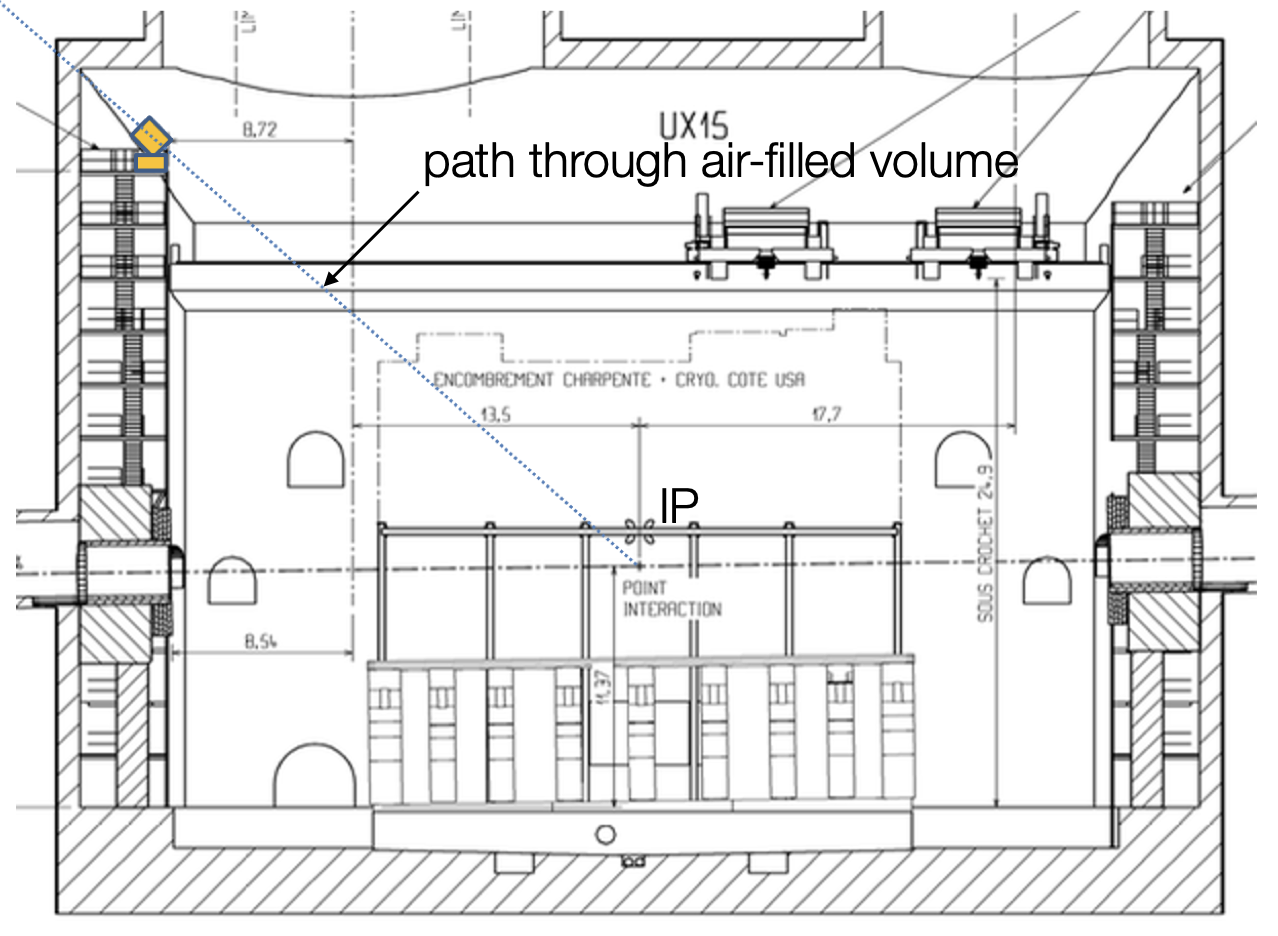}
\caption{The location of the \proanubis detector (yellow) within the ATLAS cavern.} 
\label{fig:proAnubLocation}
\end{figure}

The \proanubis detector was installed at the top of the scaffolding on the ‘A' side of the ATLAS cavern, and is rotated 45$^{\circ}$ with respect to the vertical such that the normal vector of the detector planes points away from the ATLAS interaction point\footnote{The local coordinate system of \proanubis is centred at the bottom left corner of the lowest RPC detector in the triplet RPC module when seen from the other RPC modules. The $x$ axis is aligned with the long side of the RPC detector, the $y$ axis with its short side, and the $z$ axis is perpendicular to the detector plane, forming a right-handed coordinate system. The line of sight from \proanubis to the IP approximately coincides with the $-z$ axis.}. A schematic of the detector is presented in Figure~\ref{fig:proAnubLayout}, along with its positioning in the ATLAS cavern in Figure~\ref{fig:proAnubLocation}. 
The setup is oriented such that these long and short directions approximately correspond to the ATLAS $\phi$ and $\eta$ coordinates, respectively; consequently, the long side is referred to as the $\phi$ side and the short side as the $\eta$ side, following the conventions outlined in Ref.~\cite{ANUBIS:2025inn}. 
The signal pickup strip pitch on the $\phi$ side is 2.63~cm, while on the $\eta$ side the strip pitch is 2.56~cm wide.

While \proanubis uses the signal from the ATLAS Central Trigger Processor~\cite{Butterworth:433830, Butterworth:2004nla} as a clock frequency for its data acquisition system, 
it operates independently from ATLAS otherwise. 
The detector employs a hardware-based triggering system requiring coincident hits across a configurable number of $\eta$ RPC planes within a 60~ns coincidence window.
Due to the high hit detection efficiency of $>95\%$ the $\phi$ RPC planes are not used in the trigger decision.

When a trigger is fired, Time-to-Digital Converters (TDCs) record, for each strip, the times at which the input signal crosses above and below a predefined threshold voltage within a window extending from 250~ns before to 750~ns after the arrival of the triggering signal.  
In addition, the arrival times of all trigger signals and of each bunch-counter reset (BCR) signal are continuously recorded on dedicated TDC channels. 
The BCR is a synchronisation signal distributed by the LHC timing system that marks the start of each LHC orbit and resets the bunch-crossing counter, thereby providing an absolute timing reference~\cite{Butterworth:433830}. 
This information enables the timing of each event to be determined relative to the LHC collisions.

\section{Data}
The \proanubis detector began recording data at the start of 2024~\cite{Shah:2024fpl} and collected both beam-on and cosmic data over the subsequent two years. 
\proanubis was active for approximately 104~\invfb of LHC collisions in 2024 and 73~\invfb in 2025. 
For nominal \proanubis data-taking, a trigger threshold corresponding to the requirement of four out of six RPC $\eta$ planes was used.
During LHC operation with $\it{pp}$ collisions at $\sqrt{s}$~=~13.6 TeV, the \proanubis trigger rate typically ranges from 3 to 6~kHz, depending on the applied trigger thresholds and the prevailing beam conditions.

\begin{figure}[ht]
\centering
\includegraphics[width=0.45\textwidth]{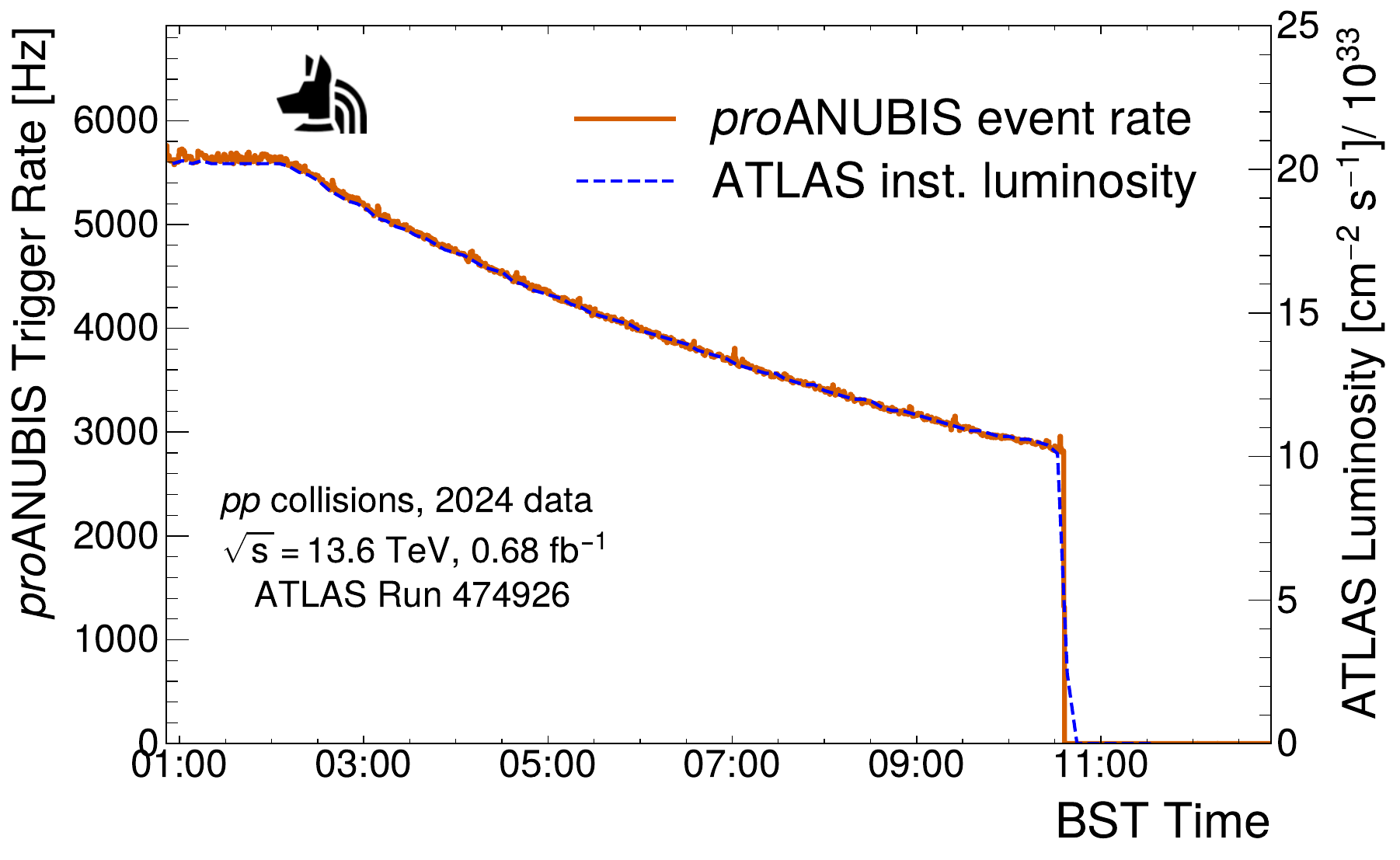}
\caption{The trigger rate in \proanubis and the ATLAS instantaneous luminosity in $pp$ collisions at $\sqrt{s}$~=~13.6 TeV for Run 474926 as a function of British Summer Time (BST).} 
\label{fig:trigRate}
\end{figure}

The \proanubis trigger rate is compared to the instantaneous luminosity measured by ATLAS for a representative LHC Run in Figure~\ref{fig:trigRate}, illustrating the strong correlation between the two. 
In absence of LHC collisions, \proanubis\ records cosmic-ray data at a rate of approximately 1~Hz.

\begin{figure}[ht]
\centering
\includegraphics[width=0.45\textwidth]{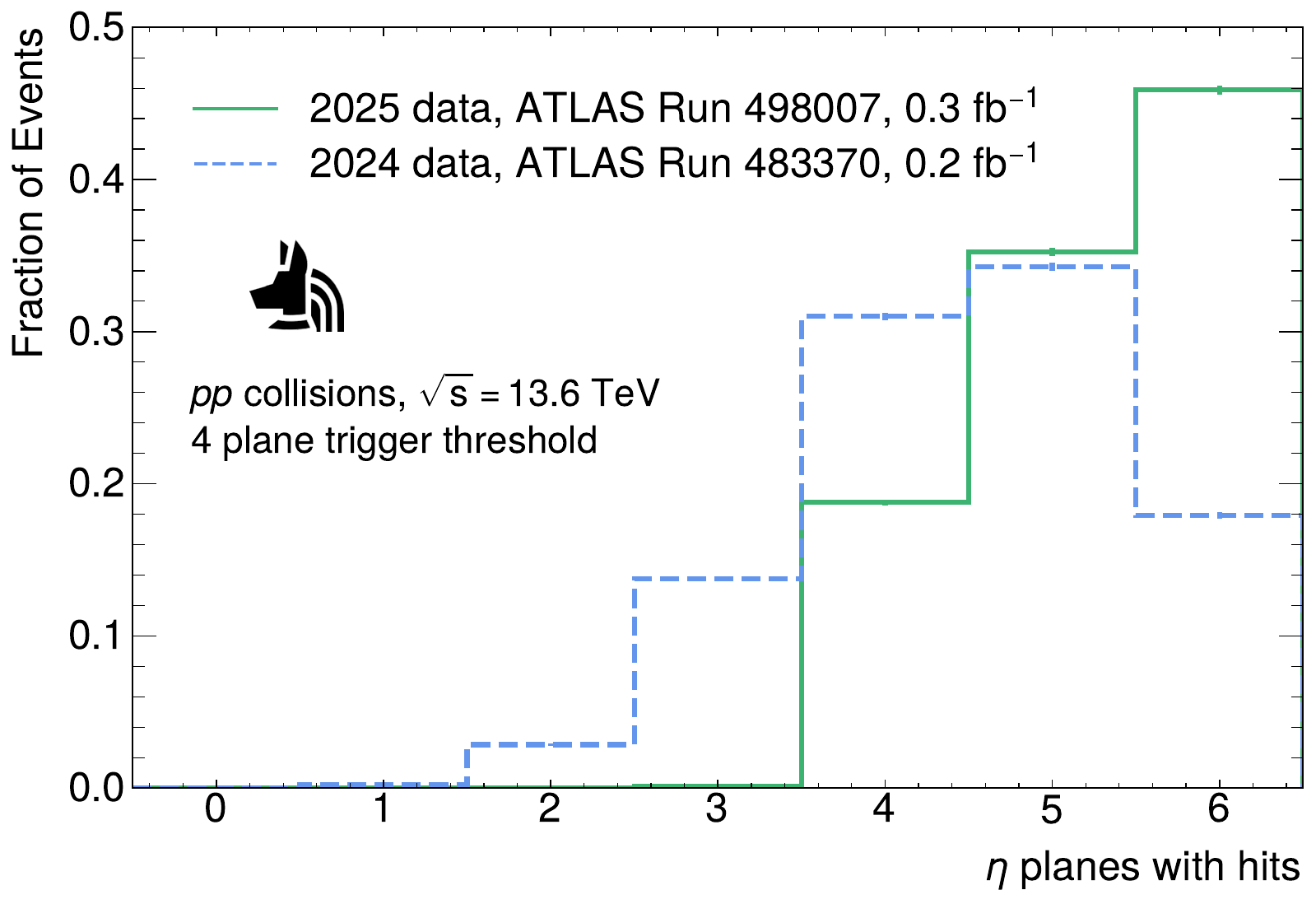}
\caption{
The number of $\eta$ RPC planes that recorded hits per event during LHC collisions with the original DAQ system in an example 2024 run compared to the upgraded DAQ for an example 2025 run with similar data-taking conditions.
} 
\label{fig:corruptEx}
\end{figure}

During the year-end technical stop at the end of 2024, the \proanubis data acquisition (DAQ) system  was upgraded, replacing the CAEN V767 TDCs with the higher-resolution model CAEN V1190 A, improving the instrumental time resolution from 800~ps to 100~ps and enhancing the overall data reliability. 
Using the original DAQ system, several issues with the data collection from the TDCs were observed, including missing data words, bit flips, and duplicated data words. 
No such issues were observed with the upgraded DAQ. 
An example of this is shown in Figure~\ref{fig:corruptEx}, which presents the distribution of the number of $\eta$ planes with recorded hits per event for comparable data-taking conditions using the 2024 and 2025 detector configurations.  

The 2024 DAQ system has significantly fewer planes registering hits compared to the upgraded version. 
This is due to cases where the hits were recorded in incorrect channels due to bit flips, or failed to be recorded, e.g., due to missing data words or failing timing requirements. 
Consequently, a significant fraction of events recorded prior to the upgrade have fewer than four $\eta$ planes with hits required by the trigger logic, indicating that a fraction of the signals contributing to the trigger decision were not recorded correctly by the DAQ.
A small fraction of events from the 2025 data are also seen with fewer than four $\eta$ planes, caused by rare residual errors in the DAQ that cause TDC information from different events to be mixed together, which are excluded from further data analysis.

\section{Event Reconstruction} \label{Sec:EventReco}
To fully assess the performance of \proanubis, dedicated algorithms have been developed to reconstruct tracks and ultimately vertices from the recorded hits. 
The reconstruction proceeds through several stages. 
First, hits are filtered to suppress electronic noise, after which they are grouped into one-dimensional (1D) clusters independently in the $\eta$ and $\phi$ readout directions of each RPC plane. The 1D clusters are then paired together to form two-dimensional (2D) clusters based on the time-coincidence criteria between the hits. 
The resulting 2D clusters are fit to straight-line tracks using singular value decomposition (SVD)~\cite{GolubReinsch1970}, and then the successful candidate tracks are re-fit using a least-squares-like algorithm. 
Finally, in events containing multiple reconstructed tracks, the tracks are back-propagated to identify and fit vertices.
These steps are discussed in more detail in the following.

\subsection{Hit Filtering and Clustering} 
To suppress electronic noise and out-of-time tracks produced in different bunch crossings, all hits are first required to fall within 35~ns of the trigger signal.
This timing acceptance threshold was chosen based on the observed variance in hit times relative to the trigger. By plotting the time of arrival (ToA) for all hits in the event, a peak is observed near the trigger time from hits correlated with the triggering event in addition to a flat background from uncorrelated hits due to noise or other particles. The 35~ns threshold is then chosen in order to accept $>$~99$\%$ of the peak. 
In addition, events with over 10 hits registered in any of the $\eta$ planes or over 20 hits in any of the $\phi$ planes are discarded in order to eliminate spurious signals associated with detector noise, which is mostly due to fluctuations in the output of the high voltage (HV) or low voltage (LV) power supplies.

1D clusters are formed iteratively, beginning with the strip exhibiting the earliest ToA in a given RPC plane. Directly adjacent strips are added to the cluster if their ToA lies within 2~ns of that of the seed strip, based on the observed time differences between adjacent strips in data.
The clustering procedure is then repeated using the outermost strips of the evolving cluster, recursively including neighbouring strips that satisfy the same adjacency and timing criteria. This process continues until no further adjacent strips meet the 2~ns time-coincidence requirement. 
Once a 1D cluster is completed, all constituent hits are removed from the pool of unclustered candidate hits, and the procedure is repeated using the earliest remaining hit until all hits have been assigned to 1D clusters. 
The position and time of each 1D cluster are defined by the strip with the earliest ToA within the cluster, which generally corresponds to the location the electron avalanche started from and hence the position and time of the particle that passed through the detector.

\begin{figure}[ht]
\centering
\includegraphics[width=0.45\textwidth]{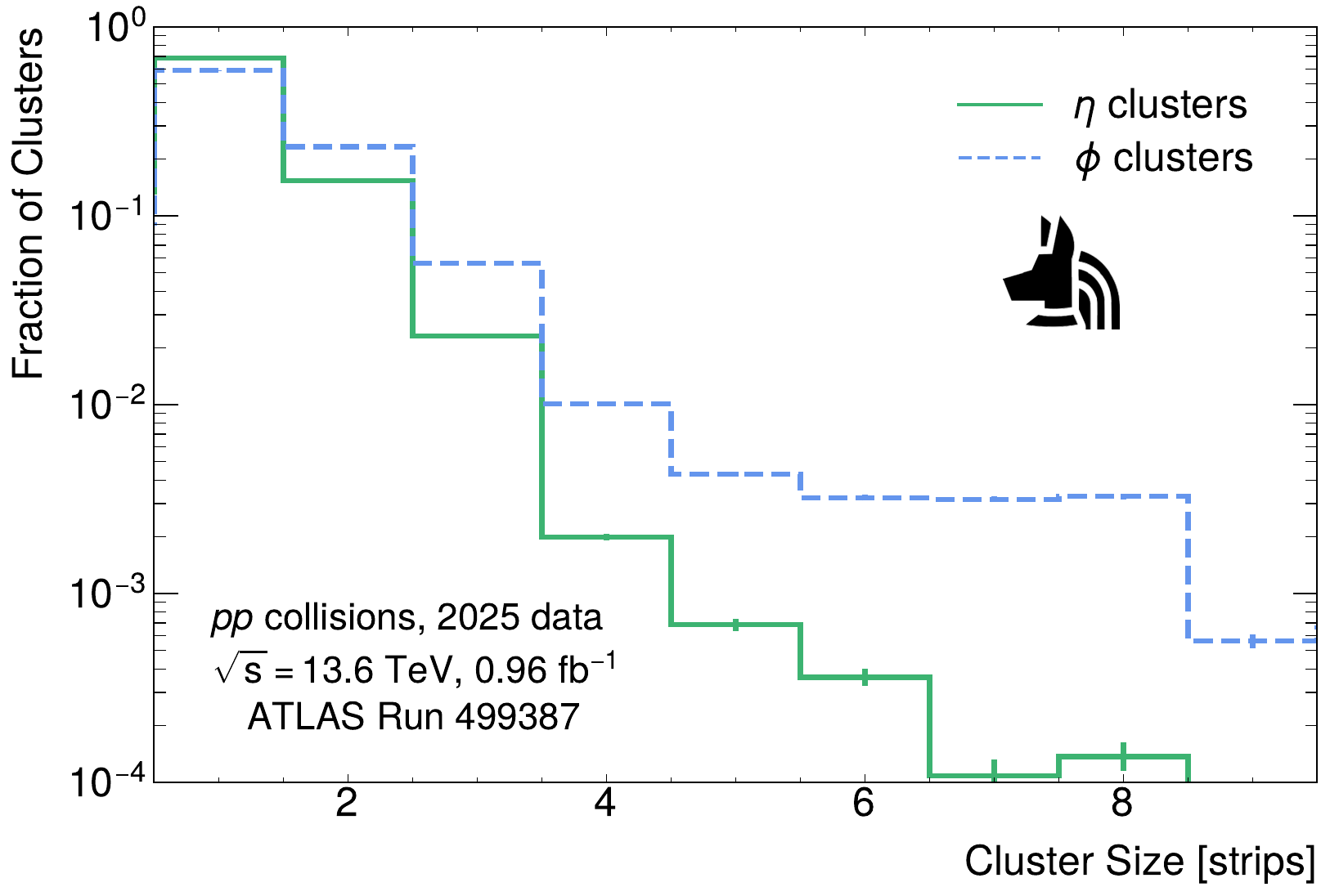}
\caption{
The 1D cluster size in the $\eta$ and $\phi$ planes in $pp$ collisions in 2025 data.
} 
\label{fig:clustSizes}
\vspace{0.5cm}
\end{figure}

\begin{figure}[ht]
\centering
\includegraphics[width=0.45\textwidth]{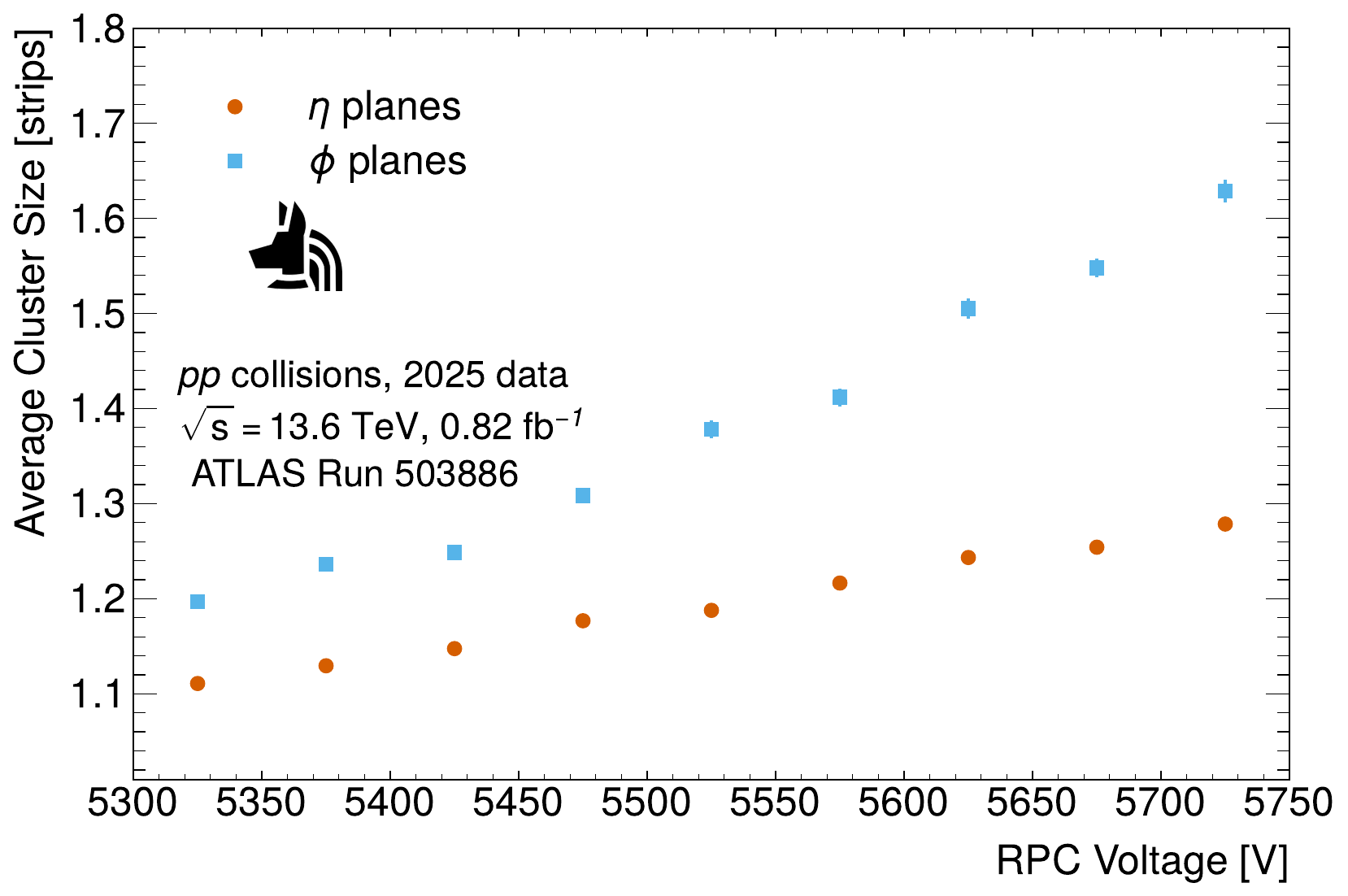}
\caption{
Average 1D cluster size in the $\eta$ and $\phi$ planes during LHC collisions as a function of applied HV in 2025 data.
} 
\label{fig:HVDepClusterSizes}
\end{figure}

The typical distribution of the 1D cluster size, i.e., the multiplicity of hits constituting a 1D cluster, is shown in Figure~\ref{fig:clustSizes} for the $\eta$ and $\phi$ RPC panels at the nominal HV operating point of 5.6 kV. 
The most probable observed cluster size for tracks in \proanubis is one strip. 
The distribution decreases significantly with increasing cluster size, with the $\phi$ plane having a higher rate of large clusters than the $\eta$ plane.
A small secondary enhancement is observed for clusters comprising eight strips, arising from the grouping scheme of the front-end electronics. 
Figure~\ref{fig:HVDepClusterSizes} shows the dependence of the average cluster size on the HV applied. 
The average cluster size is found to increase with growing HV, which is due to the increase in the typical avalanche size and in the probability of secondary ionisation.
The $\eta$ RPC panels are observed to have smaller cluster sizes than the $\phi$ panels.
This is expected to be caused by the opposite polarities of the signals induced by the electron avalanche when registered on the two sides of the gas gap.

For each RPC plane, the pairwise time difference between every 1D cluster in the $\eta$ and $\phi$ directions is evaluated and compared with the expected time difference for tracks at that $(\eta$, $\phi$) coordinate. 
The expected time differences are determined experimentally using a control sample of tracks with one cluster in each plane. 
2D clusters are then formed by accepting the $\eta$/$\phi$ pair whose time difference is closest to the expectation. 
Once a pair is accepted, all other combinations involving either of the constituent 1D clusters are discarded, and the procedure is repeated iteratively until no remaining candidate pairs have a time difference within 5~ns of the expected value.

If any given 1D cluster results in two or more candidate ($\eta$, $\phi$) pairs whose time differences from the expected value are both less than 150~ps, that cluster and the resulting candidate 2D clusters are classified as \emph{ambiguous}. 
Such cases typically arise in events with two tracks separated by an equal number of $\eta$ and $\phi$ strips. 
If ambiguous clusters are present in a given RPC plane, the set of ($\eta$, $\phi$) combinations that minimise the overall spread in time differences among all ambiguous 2D cluster candidates in the RPC plane is selected. 
This choice is motivated by the assumption that tracks originating from the same event reach the \proanubis\ detector within a narrow time window.

\begin{figure}[ht]
\centering
\includegraphics[width=0.45\textwidth]{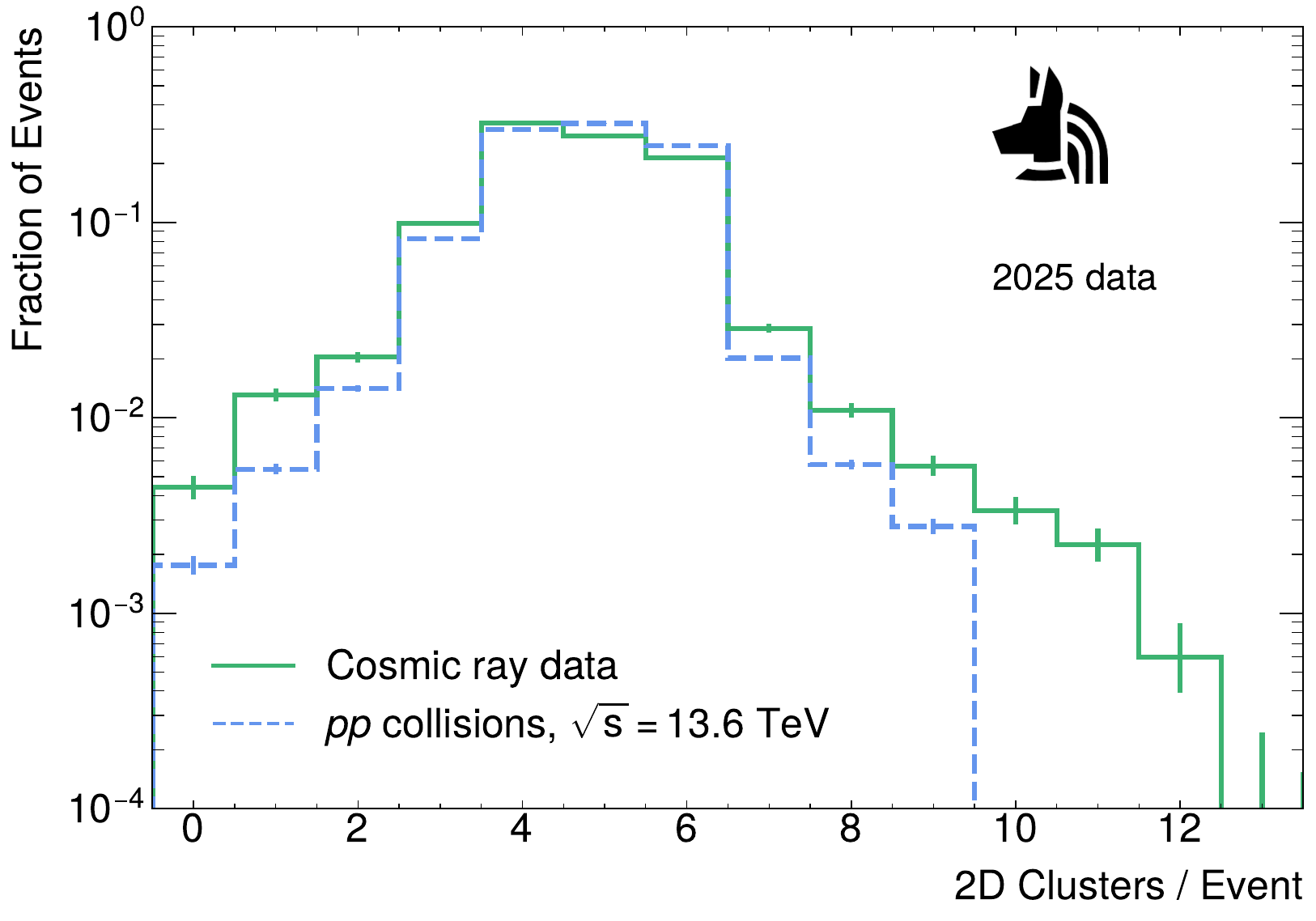}
\caption{
The number of 2D clusters per event for a sample of cosmic ray data and $pp$ collision data collected on the same day.
} 
\label{fig:2DClusters}
\end{figure}

After pairing, the time of each 2D cluster is then defined as the average of the times of the earliest constituent $\eta$ and $\phi$ strips. 
The distribution of the number of 2D clusters per event for a sample of cosmic ray data and a sample of LHC collision data is shown in Figure~\ref{fig:2DClusters}. 
In both samples, the majority of events contain between four to six 2D clusters, consistent with tracks produced by single charged particles traversing the detector.
The cosmic ray data exhibit an increased relative population of events at both low and high cluster multiplicities. 
This behaviour is attributed to the larger spread of track incidence angles and the lower trigger rate, which enhances the relative contribution of electronic noise to triggered events.
Events with cosmic ray showers may also have larger cluster multiplicities due to multiple particles traversing \proanubis. 

\subsection{Track Fitting}
The clusters are combined to form tracks using a three-stage process.
First, track `seeds' are constructed independently within each tracking layer. 
These seeds are then paired across different layers to identify candidate tracks spanning the detector volume. 
Finally, the candidate tracks are fit using a least-squares-like algorithm. 
 
In the seeding stage, clusters within each tracking layer are paired together based on their spatial proximity in order to reduce the number of possible combinations in the track fitting stage. 
Starting in the lowest RPC of each tracking layer, each cluster is used to initialise a new seed. 
Next, for each other RPC plane within the same tracking layer, candidate clusters are pre-matched to the seed by finding all clusters with physical distances to a given seed of less than five times its position resolution in both $\eta$ and $\phi$, where for simplicity the position resolution is chosen to be the strip ${\mathrm{pitch~} / \sqrt{12}}$.
Seeds are then grown by selecting the pre-matched seed-cluster pair with the smallest distance and dropping all other possible pre-matched combinations involving those clusters.
In cases where multiple candidate matches yield identical distances, the selection is resolved sequentially using the physical distance to potential clusters in the neighbouring RPC plane, followed by the smallest time difference within the given RPC plane, then finally the largest Time Over Threshold (TOT), if necessary, in order to select clusters that are likelier to fit to tracks or correspond to more significant energy deposited in the RPC. 
Any 2D clusters that are not matched to existing seeds are made into new seeds.

After seed formation, track candidates are constructed by fitting every possible combination of seeds in the Triplet and Doublet layers, considering only seeds that contain at least one 2D cluster. 
The initial candidate tracks are fitted using a SVD of all clusters associated within the selected seeds. If a viable track candidate is obtained, it is propagated into the Singlet layer. 
For each 2D cluster found within five times its position uncertainty of the extrapolated track, a new track candidate is formed by combining that cluster with the original Triplet and Doublet seeds. 
The resulting set of candidate tracks is then iteratively selected by accepting the candidate with the largest number of seeds, followed by the lowest reduced $\chi^2$ in the case of ties, removing all other candidates using seeds in the accepted track, and repeating the process until no tracks remain in the candidate pool.

Lastly, the accepted tracks are refit to fully account for the relative uncertainty of the individual cluster coordinates and to obtain uncertainties on the reconstructed track parameters. 
For each RPC plane, a $\chi^2$ significance is defined using the distance from the projected track in that plane to the location of the cluster centre relative to the uncertainty on the cluster position. 
For 1D clusters, the significance can thus be defined using only the coordinate of the 1D cluster and the corresponding track coordinate.
After refitting, only tracks with reduced $\chi^2<3$ are accepted.

\begin{figure}[ht]
\centering
\includegraphics[width=0.45\textwidth]{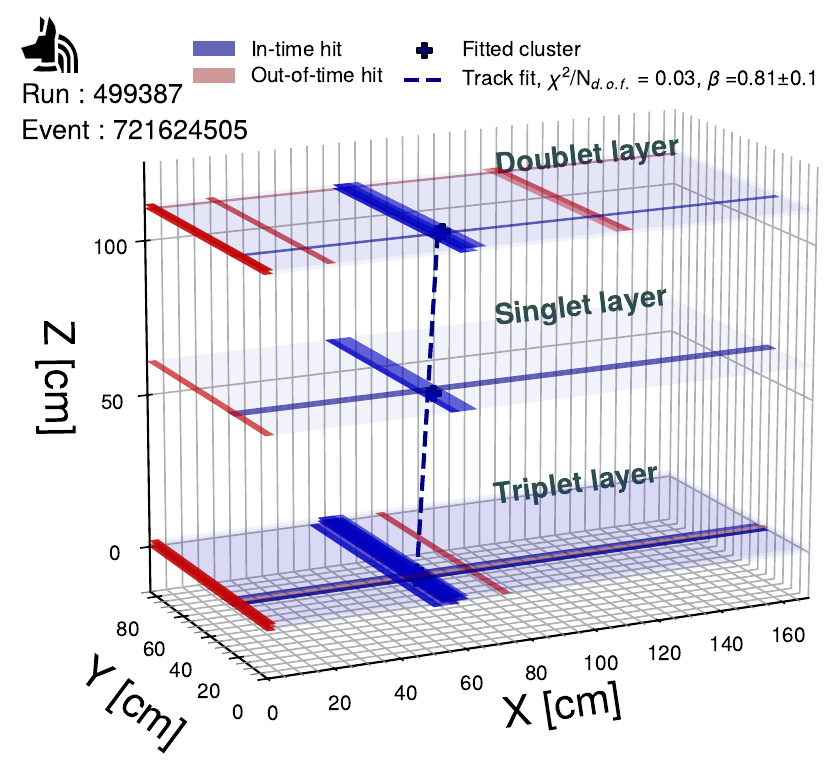}
\caption{A \proanubis event display for a single-track event in $pp$ collisions.} 
\label{fig:oneMuEventDisplay}
\end{figure}

A display of a typical \proanubis event collected in $pp$ collisions is shown in Figure~\ref{fig:oneMuEventDisplay}.
The display uses the \proanubis local coordinate system. 
The in-time hits with ToA within 35~ns of the trigger signal in the event are successfully reconstructed as a track with direction and velocity consistent with a high-energy particle originating from the ATLAS IP. 
The out-of-time hits with ToA $>$35~ns away from the trigger signal are shown in red. The out-of-time hits observed on the first channel in the $\phi$ direction over several planes are likely due to correlated noise on the low-voltage inputs, which are positioned along that edge of the RPC planes.
In the ATLAS data for this event, a muon with trajectory consistent with proANUBIS is also observed.

\begin{figure}[ht]
\centering
\includegraphics[width=0.45\textwidth]{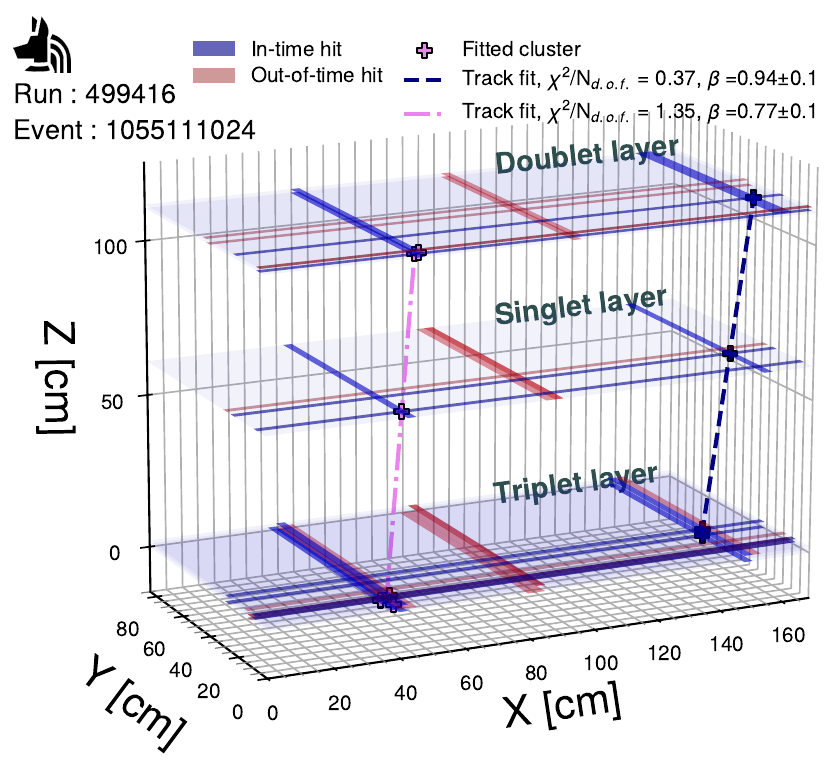}
\caption{A \proanubis event display for a two-track event in $pp$ collisions.} 
\label{fig:twoMuEventDisplay}
\end{figure}

A display of representative \proanubis event with two reconstructed tracks collected in $pp$ collisions is shown in Figure~\ref{fig:twoMuEventDisplay}. 
In this event, the pairing requirements used in the cluster formation consistently select the same corners in each RPC plane, and two tracks are reconstructed. The direction and velocities of both tracks are consistent with particles produced at the IP. 
In the ATLAS data for this event, two oppositely-charged muons are observed which both have trajectories consistent with particles passing through \proanubis.

\section{RPC Efficiency}
The efficiency of each RPC plane is evaluated by excluding that plane from the track reconstruction, which otherwise follows the procedure described in Section~\ref{Sec:EventReco}.
Additionally, the non-excluded planes are required to pass the \proanubis trigger to avoid bias from the excluded plane.
The efficiency is then measured by determining the fraction of events where a cluster is observed in geometrical vicinity of the expected track position in the RPC plane under study.

\begin{figure}[ht]
\centering
\includegraphics[width=0.45\textwidth]{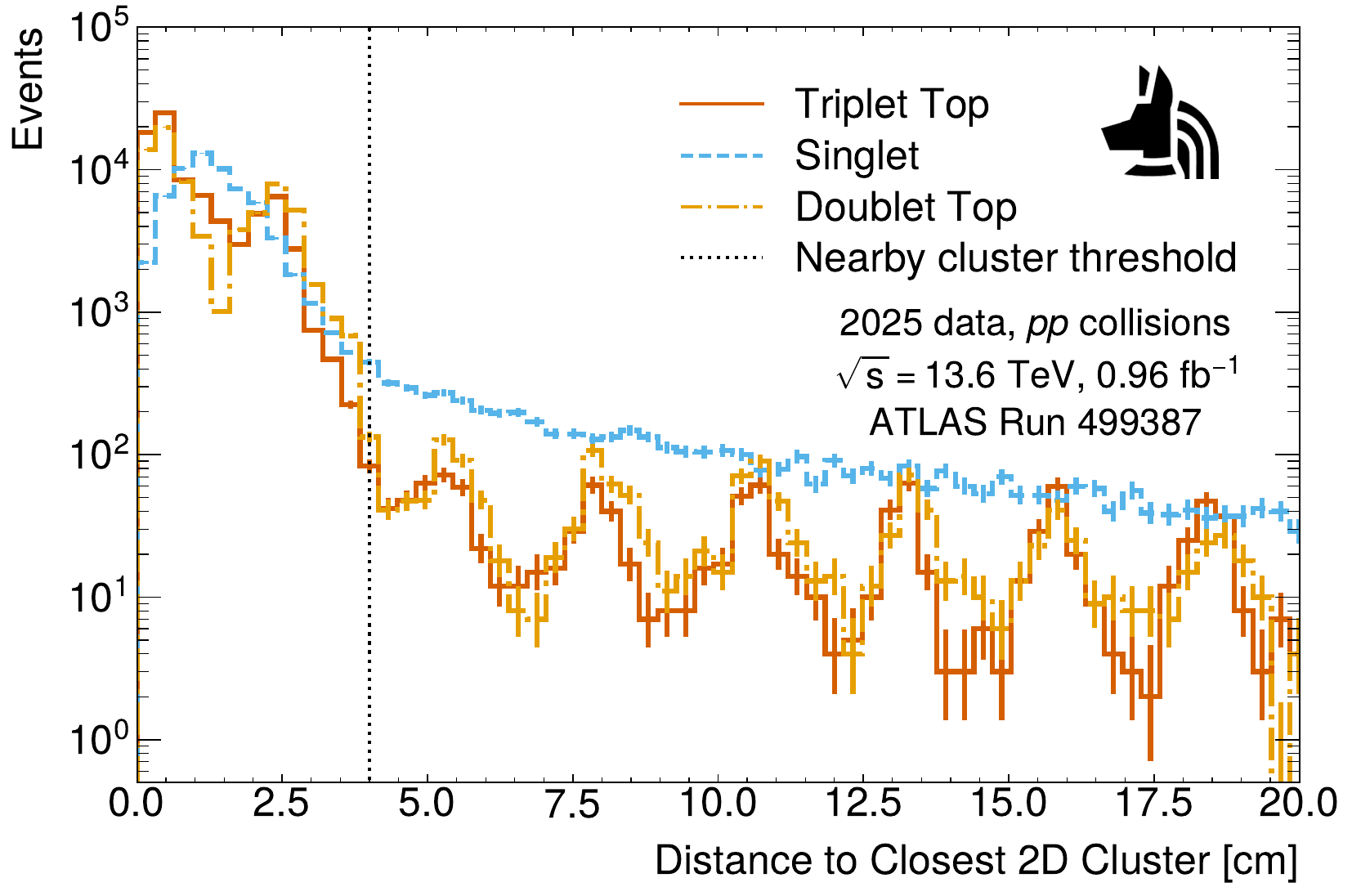}
\caption{
The distance between projected track position and the nearest 2D cluster in three example RPC planes. The threshold for associating the 2D cluster with a track for the purpose of the efficiency measurement is indicated by the dotted vertical line.
} 
\label{fig:2Ddistance}
\end{figure}

When measuring the efficiency of detecting 1D clusters, the $\eta$ and $\phi$ readout planes are considered separately, and a distance threshold of 4~cm is applied to determine whether a cluster is in the vicinity of the projected track. 
For 2D clusters, both $\eta$ and $\phi$ readout planes are considered and the clusters are required to be within 4~cm from the projected track position in the RPC plane under study. 
The distribution of the 2D cluster distance is shown in Figure~\ref{fig:2Ddistance} for three representative RPC planes.
The RPC planes in the Triplet and Doublet display an exponentially falling trend as a function of 2D cluster distance up to $\approx$4~cm, followed by an approximately flat shoulder. In addition to the general trend, periodic oscillations with a characteristic distance of $\approx$2.5~cm are observed due to the granularity of the RPC readout strips. 
For the Singlet, the distribution follows the same general trend as the Triplet and Doublet layers, but does not display any periodic oscillations as there is no directly adjacent RPC plane to strongly constrain the track fit to the strip centers.

\begin{figure}[ht]
\centering
\includegraphics[width=0.45\textwidth]{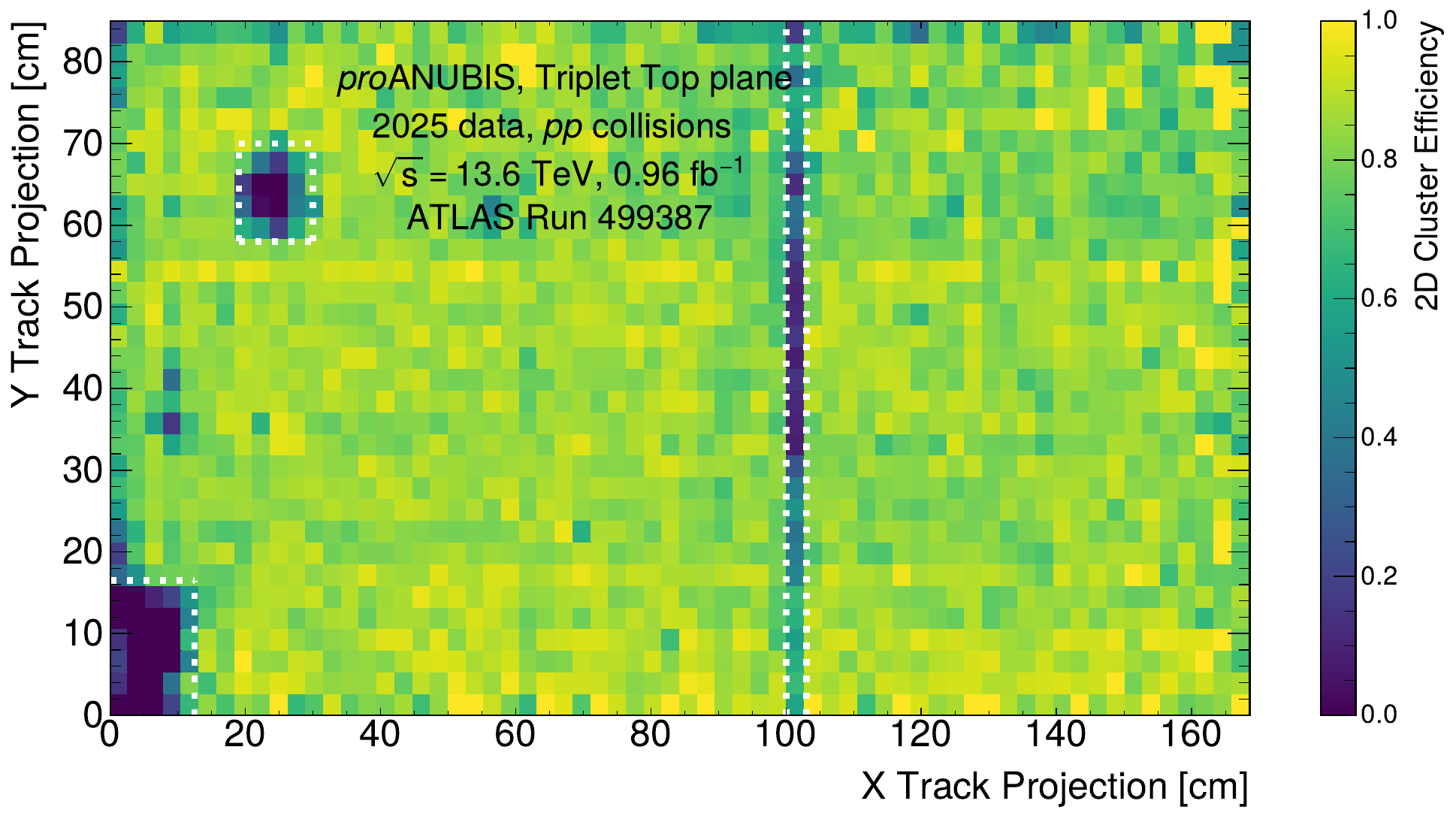}
\caption{
The RPC hit efficiency in the Triplet Top plane as a function of the $X$ and $Y$ position of the projected track, with the low-efficiency regions outlined by a dotted white line.} 
\label{fig:2Deff}
\end{figure}

To account for inactive strips or localised regions on the RPC with a systematically lower efficiency, the RPC efficiency is first measured as a function of the projected track position across each plane. 
Regions exhibiting significantly lower efficiency are excluded from the global efficiency measurement.
These regions are typically caused by issues with the electronic readout and localized flaws in the gas gap of the chamber.
This approach ensures that measurements of the efficiency dependence on the applied HV are not biased by variations in the active area of individual planes. 
An example efficiency map for the Triplet Top plane alongside regions excluded from the efficiency measurement is shown in Figure~\ref{fig:2Deff}.

\begin{figure*}[ht]
\centering
\subfloat[]{\includegraphics[width=0.5\textwidth]{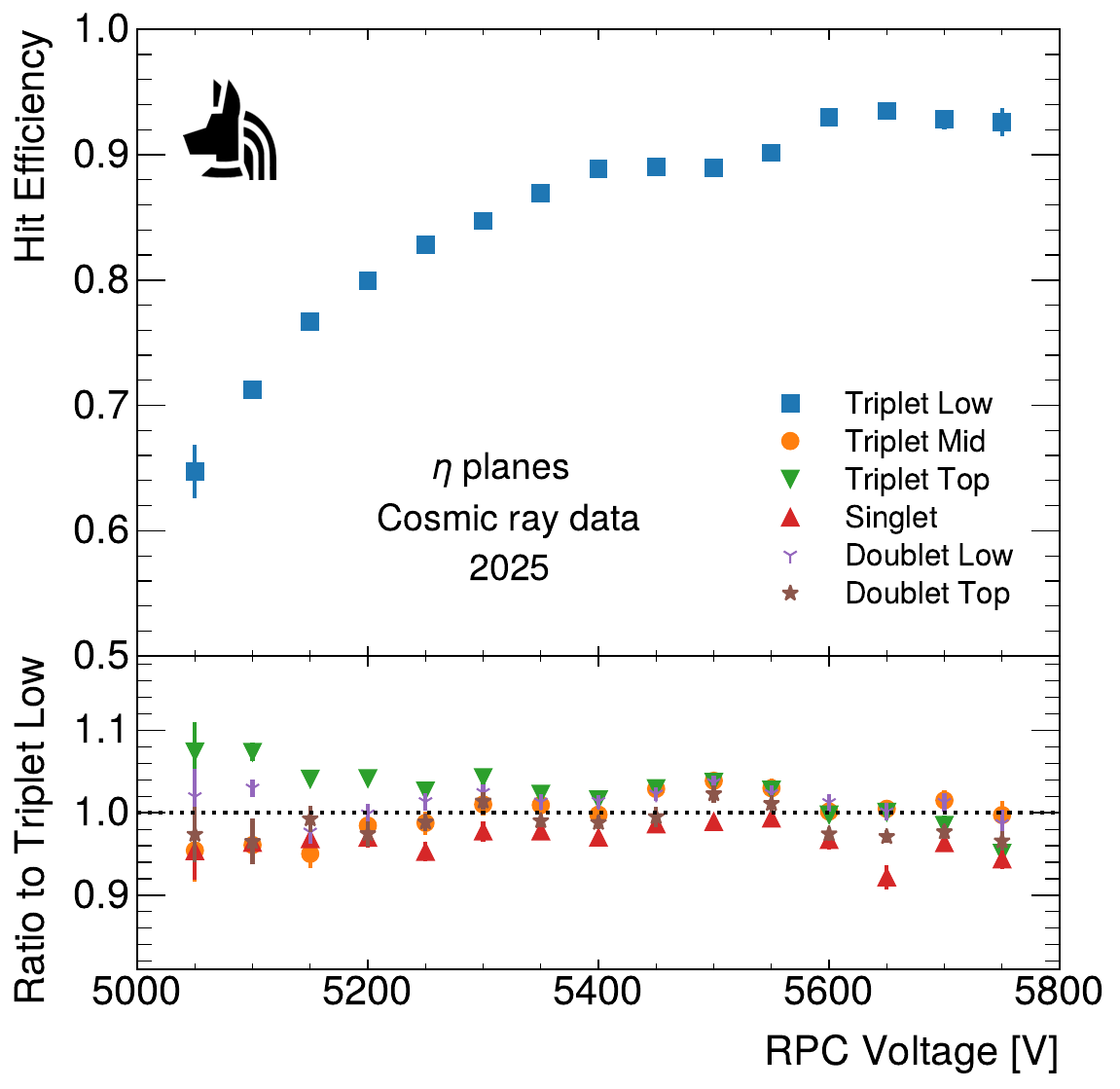}}
\subfloat[]{\includegraphics[width=0.5\textwidth]{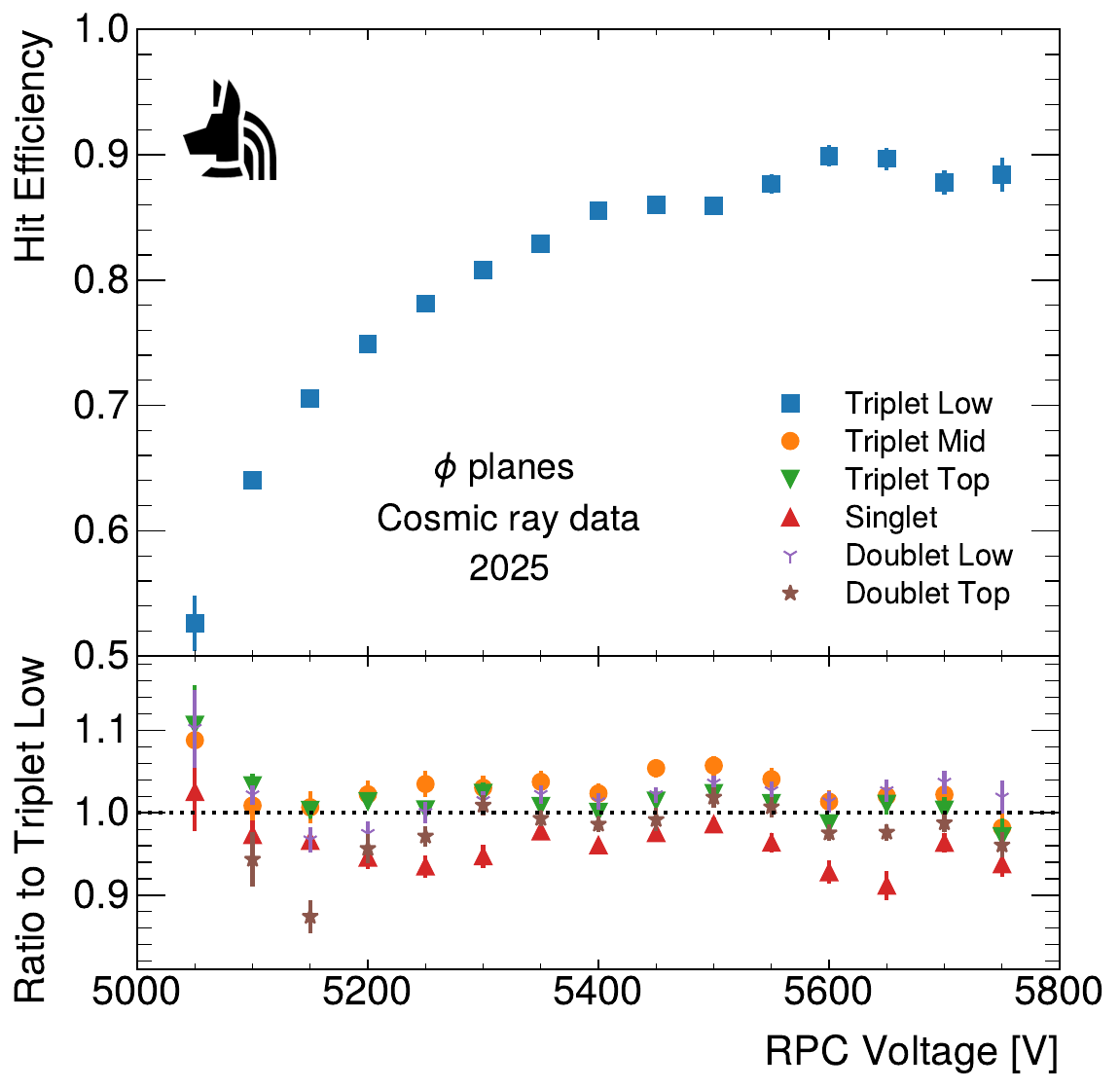}}
\caption{The RPC efficiency measured in cosmic ray data as a function of applied HV to the RPC plane under study. 
The efficiency of the $\eta$ plane is shown in panel~(a), and the $\phi$ plane in panel~(b).} 
\label{fig:stripEff}
\end{figure*}

The resulting 1D cluster efficiencies of each RPC in \proanubis are shown as a function of the applied HV, separately for the $\eta$ and $\phi$ readout strips in Figure~\ref{fig:stripEff}. 
All RPCs reach efficiency plateaus exceeding 85$\%$ in both directions, corresponding to an overall efficiency of $>$~99\% for the OR combination of a three-plane tracking layer. 
The $\eta$ readout planes are observed to be consistently more efficient than the $\phi$ planes. 
This feature is attributed to the polarity of the gas gap: the $\eta$ strips are on the positive side of the gas gap and hence closer to the maximum electron avalanche.

\begin{figure}[ht]
\centering
\includegraphics[width=0.45\textwidth]{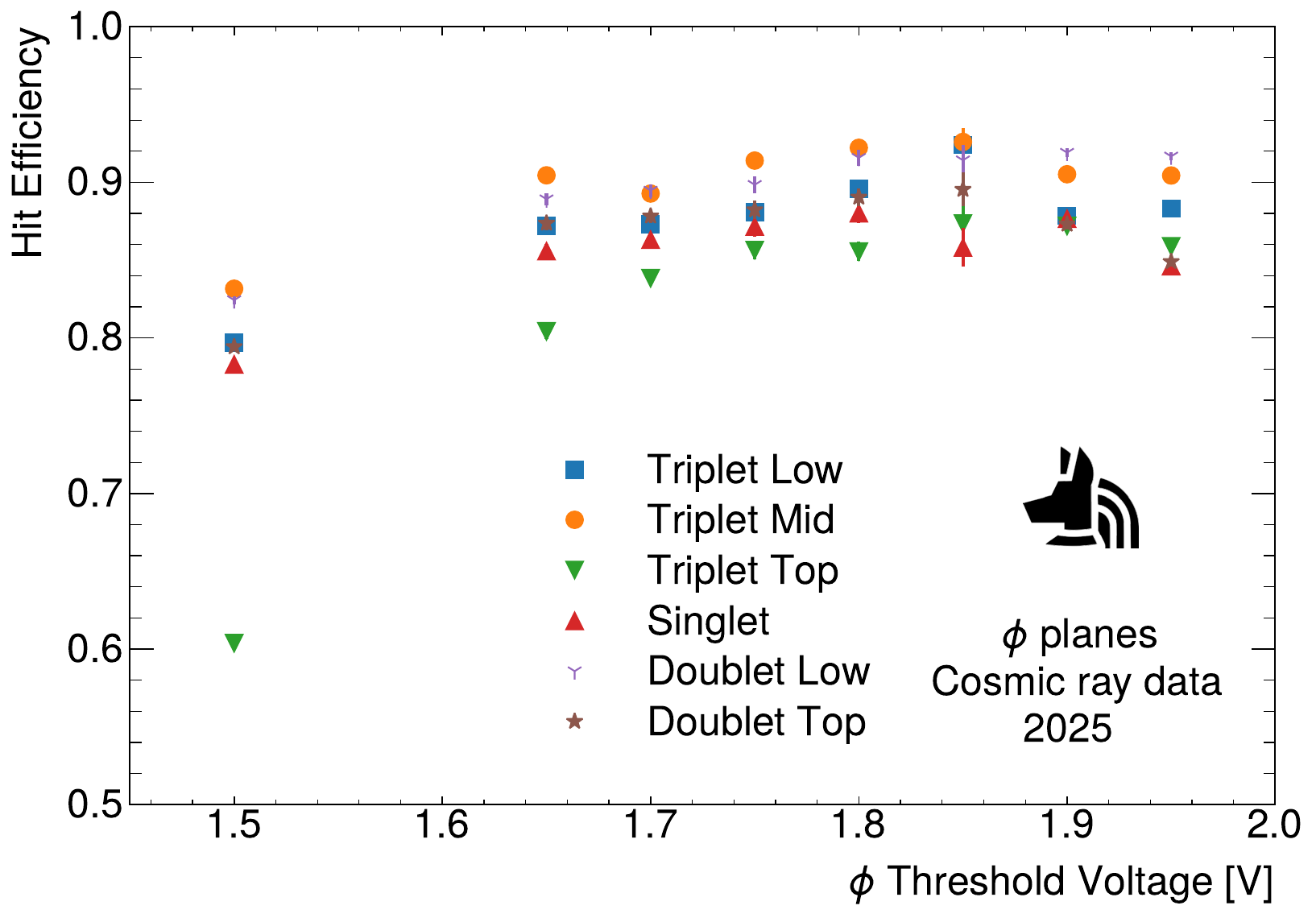}
\caption{The RPC hit efficiency in all $\phi$ planes as a function of the applied threshold voltage.} 
\label{fig:ThreshEff}
\end{figure}

The efficiency of each RPC plane as a function of the $\phi$ threshold voltage is measured using an identical method of fitting tracks with individual RPC planes excluded. 
This threshold voltage sets the reference for the discriminator to register a signal on a given strip, with higher threshold voltage corresponding to a smaller required  amplitude of the input signal. 
This dependence is shown in Figure~\ref{fig:ThreshEff}, where the larger input signal required at low threshold voltage results in correspondingly lower efficiency. 
During the data collection for this study, the Triplet Top plane was operated at a reduced high voltage of 5.4~kV compared to 5.6~kV for the other planes, likely producing the lower overall efficiency and the larger decrease in efficiency at low threshold observed.

\section{Calibration}
\label{sec:Calibration}
In order to accurately pair hits, determine track timing, and measure the cluster position several systematic effects must be calibrated for. 
These include the finite propagation speed of electrical signals along the RPC strips, channel-dependent timing offsets arising from variations in signal paths that are driven by electronic component fabrication tolerances, and ToA variations associated with differences in the avalanche signal amplitude.

\subsection{Strip-level calibrations}
\label{Sec:sigProp}
Electrical signals induced on the RPC readout strips must first propagate to the detector edge before being processed by the front-end (FE) electronics, introducing a $\phi$-dependent delay to the registered $\eta$ hit times and an $\eta$-dependent delay to the $\phi$ hit times. 
This signal propagation speed is determined using a combined fit of the average $\eta$ - $\phi$ time difference for all pairwise combinations of strips in each plane.
Only those events where the corresponding strips record the earliest hit times in a given RPC plane are considered. 
By performing a combined fit across all RPC layers, with the propagation velocities in the $\eta$ and $\phi$ directions constrained to be identical for each plane, the effective signal propagation speed is measured to be approximately $0.6\,c$.

\begin{figure}[ht]
\centering
\includegraphics[width=0.45\textwidth]{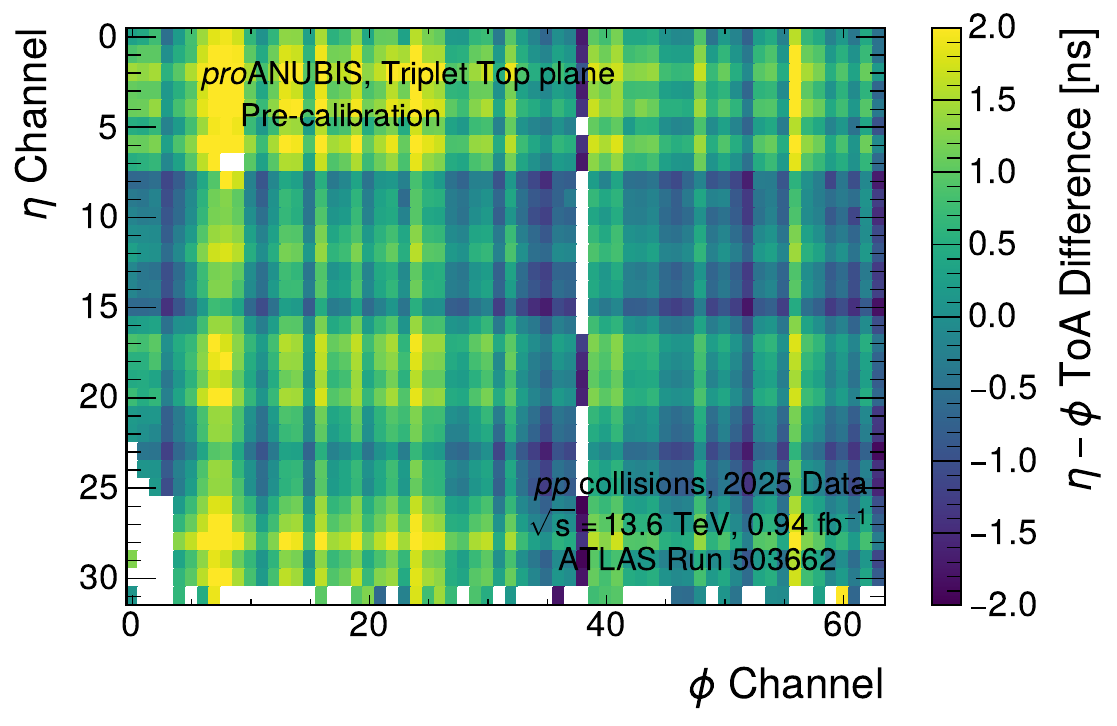}
\caption{Average ToA differences for each pairwise combination of $\eta$ and $\phi$ channels in the Triplet Top layer with the signal propagation time and overall $\eta-\phi$ scale removed.
The white cells correspond to $\eta$ - $\phi$ ToA differences that have no entries or are outside of the $[-2,2]$~ns time window.
} 
\label{fig:preCalibPlane}
\end{figure}

For 2D clusters, the $\eta$ and $\phi$ times are thus corrected for the signal propagation by subtracting the signal speed multiplied by the distance to the detector edge along the other coordinate from each ToA. 
For 1D clusters, the signal propagation delay is instead estimated using the coordinates of the track fit projected into the corresponding RPC plane. 
Following the signal propagation correction, the average $\eta$ - $\phi$ time difference for each pair of strips is re-calculated with the signal propagation subtracted. 
The resulting $\eta$ - $\phi$ ToA differences are shown in Figure~\ref{fig:preCalibPlane} for the Triplet Top RPC layer as an example.

To correct for residual strip-dependent offsets arising from variations in electrical connections, FE electronics, and TDC channels, a per-strip timing calibration is applied. 
For each $\eta$ channel the median time difference for all $\phi$ channels is computed. 
For channel zero, this time difference is taken as a constant offset.
Beginning with channel one, each strip is then assigned a constant offset such that its median time difference is equal to that of the preceding strip. 
In order to remove the impact of inactive strips, if the preceding strip has a mean hit occupancy of $<$~10\% of the mean of all strips in a given RPC plane the nearest preceding strip with occupancy above the 10\% threshold is used instead. 

\begin{figure}[ht]
\centering
\includegraphics[width=0.45\textwidth]{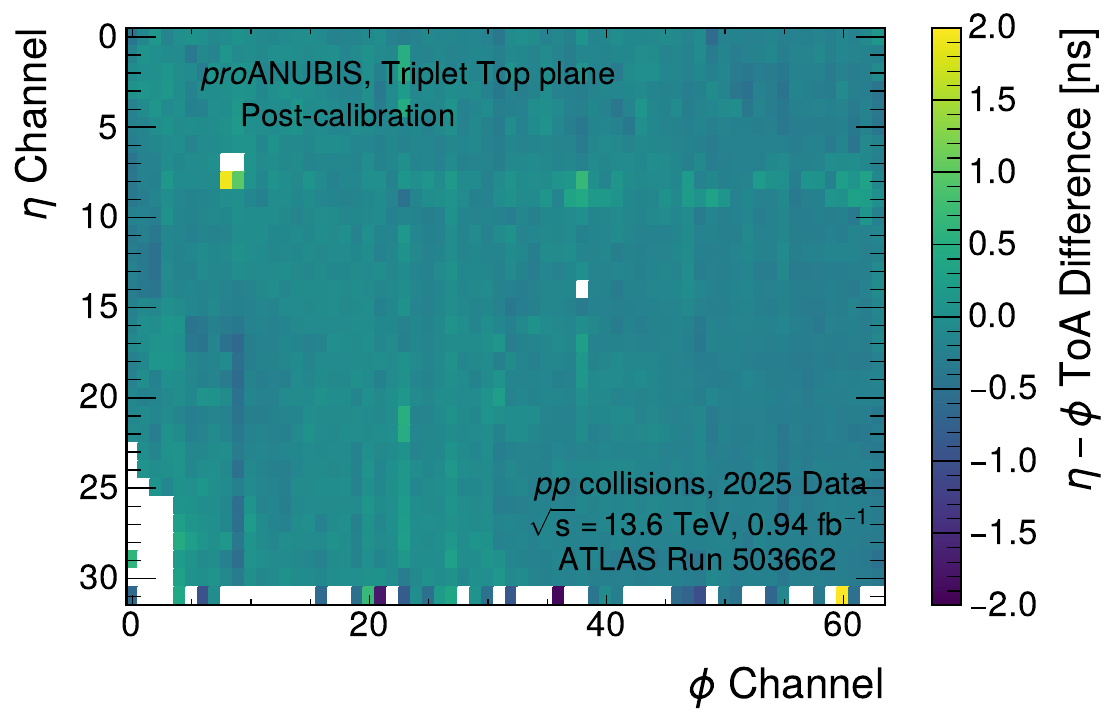}
\caption{Average ToA differences for each pairwise combination of $\eta$ and $\phi$ channels in the Triplet Top layer with all systematic strip calibrations applied.
The white cells correspond to $\eta$ - $\phi$ ToA differences that have no entries or are outside of the $[-2,2]$~ns time window.
} 
\label{fig:postCalibPlane}
\end{figure}

The process is then repeated for the $\phi$ direction. 
The $\eta$ - $\phi$ ToA differences after the strip-level calibration are shown in Figure~\ref{fig:postCalibPlane} for the Triplet Top RPC layer as an example. 
The pre-calibration data show clear systematic variations of a couple ns, which are effectively removed by the strip-level timing calibrations.

\subsection{RPC Plane Offsets}
Within each RPC plane, the $\eta$ strip times exhibit additional systematic delay with respect to the $\phi$ strip times, typically around 12~ns, due to the electronics and cabling used for the trigger decision. 
To correct for this plane-level offset, the average time difference for each $\eta$ and $\phi$ pair is computed for each plane after applying the signal propagation and strip-level timing calibrations. 
A constant offset equal to this average time difference in a given RPC plane is then subtracted from the $\eta$ times in that plane, thereby removing the residual plane-dependent timing bias.

Similarly, each RPC plane may have an overall time offset relative to the other layers due to small variations in the readout components associated with that plane. 
These inter-plane offsets are calibrated using samples of tracks from LHC collisions as follows.
All tracks used are required to be consistent with originating from the ATLAS IP in order to reduce backgrounds from cosmic muons and lower-velocity collision products. 
Starting from the first plane (Triplet Low), the time difference between 2D clusters in that RPC plane and in the plane immediately above it is evaluated for tracks with clusters in both planes.
This difference is then compared to the expected time-of-flight (ToF) for $\beta=$~1. 
The distribution of time differences is fitted with a Gaussian function, and the fitted mean is taken as a constant offset for all times within the higher plane. 
The process is then iteratively repeated for every pair of RPC planes to derive plane-to-plane corrections for the entire detector. 

\begin{figure}[ht]
\centering
\includegraphics[width=0.45\textwidth]{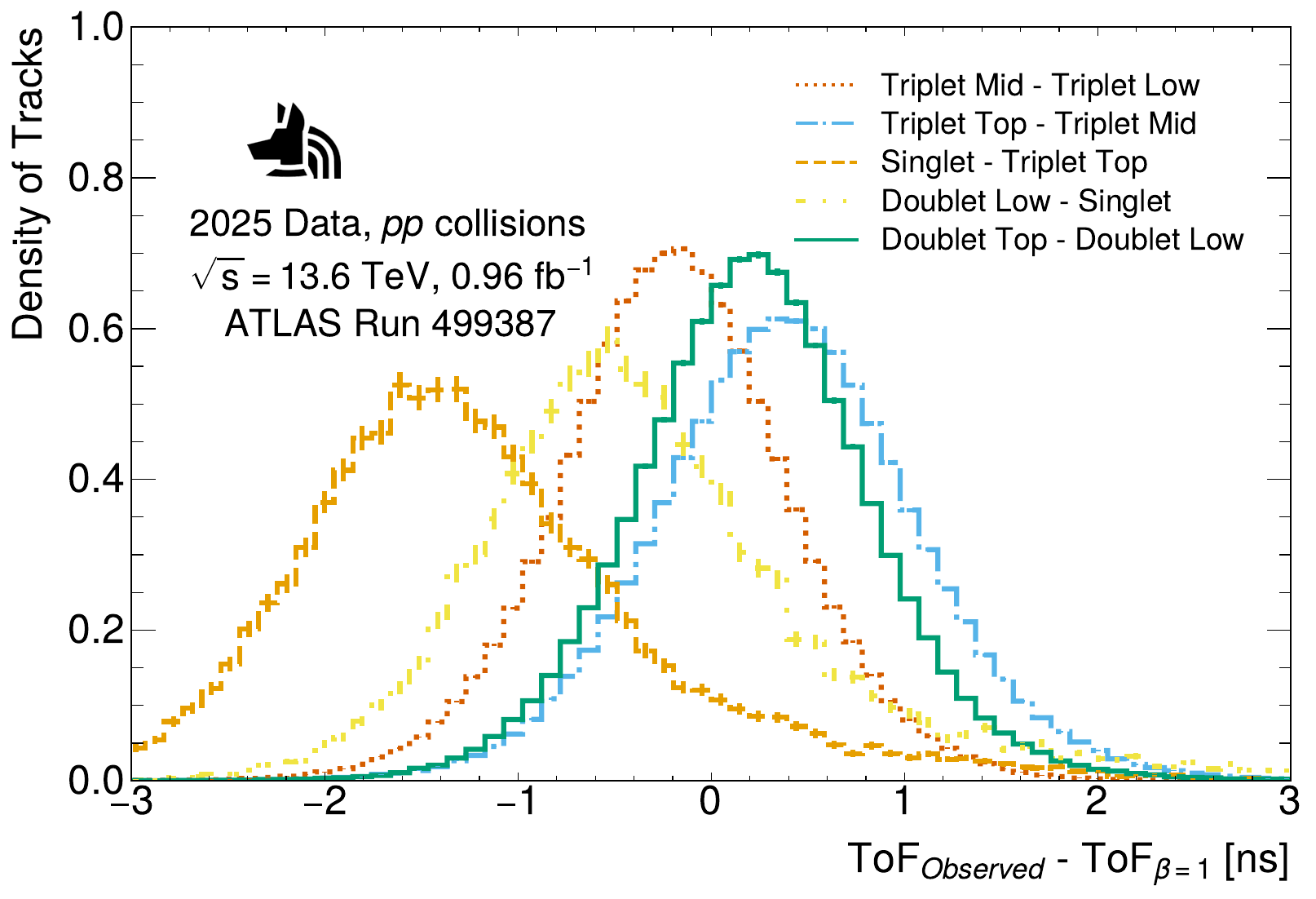}
\caption{Observed time-of-flight for RPC planes with adjacent indices, relative to the expected time difference for tracks with $\beta$~=~1.} 
\label{fig:planeOffsetFits}
\end{figure}

The ToF distributions for each pair are shown in Figure~\ref{fig:planeOffsetFits}. The extracted plane offsets are on the order of $\mathcal{O}$(0.1)~ns for planes within the same layer, and $\mathcal{O}$(1)~ns across layer boundaries. The two pairs involving the Singlet layer exhibit significantly wider distributions due to contributions from slow tracks and the relatively longer path length from the Singlet to its nearest adjacent planes.

\subsection{Time Walk}
An additional correction is applied to the ToA of each hit due to the correlation between ToA and TOT in the RPC readout strips. 
The ToA is measured as the time when the input voltage surpasses the threshold voltage. 
Therefore, larger signals with identical truth arrival times will pass the threshold sooner and hence appear to arrive earlier than lower signals. 
This amplitude-dependent bias, commonly referred to as ``time walk'', is corrected for via a two-stage calibration procedure described in the following.

First, for each pair of RPC planes with adjacent indices, samples of events containing tracks with clusters reconstructed in both planes are selected, similarly to the inter-plane offset calibration.
A 2D distribution is then constructed using the TOT measured in the lower plane (TOT$_0$) and the difference in ToA between the two planes relative to the expected ToF. 
At each value of TOT$_0$, the time difference distribution is fit to a Gaussian function, and the fitted mean values of the Gaussian function are saved as a TOT-dependent calibration for a given plane. 

Next, a 2D distribution is created using the TOT of the higher plane (TOT$_1$) and the time difference between the two planes after correcting for TOT$_0$, which is similarly fit for each value of TOT$_1$ and the Gaussian centres saved at each TOT$_1$. The final time difference between the two planes is then obtained by subtracting both the TOT$_0$- and TOT$_1$-dependent corrections from the measured time difference.

\begin{figure}[ht]
\centering
\includegraphics[width=0.45\textwidth]{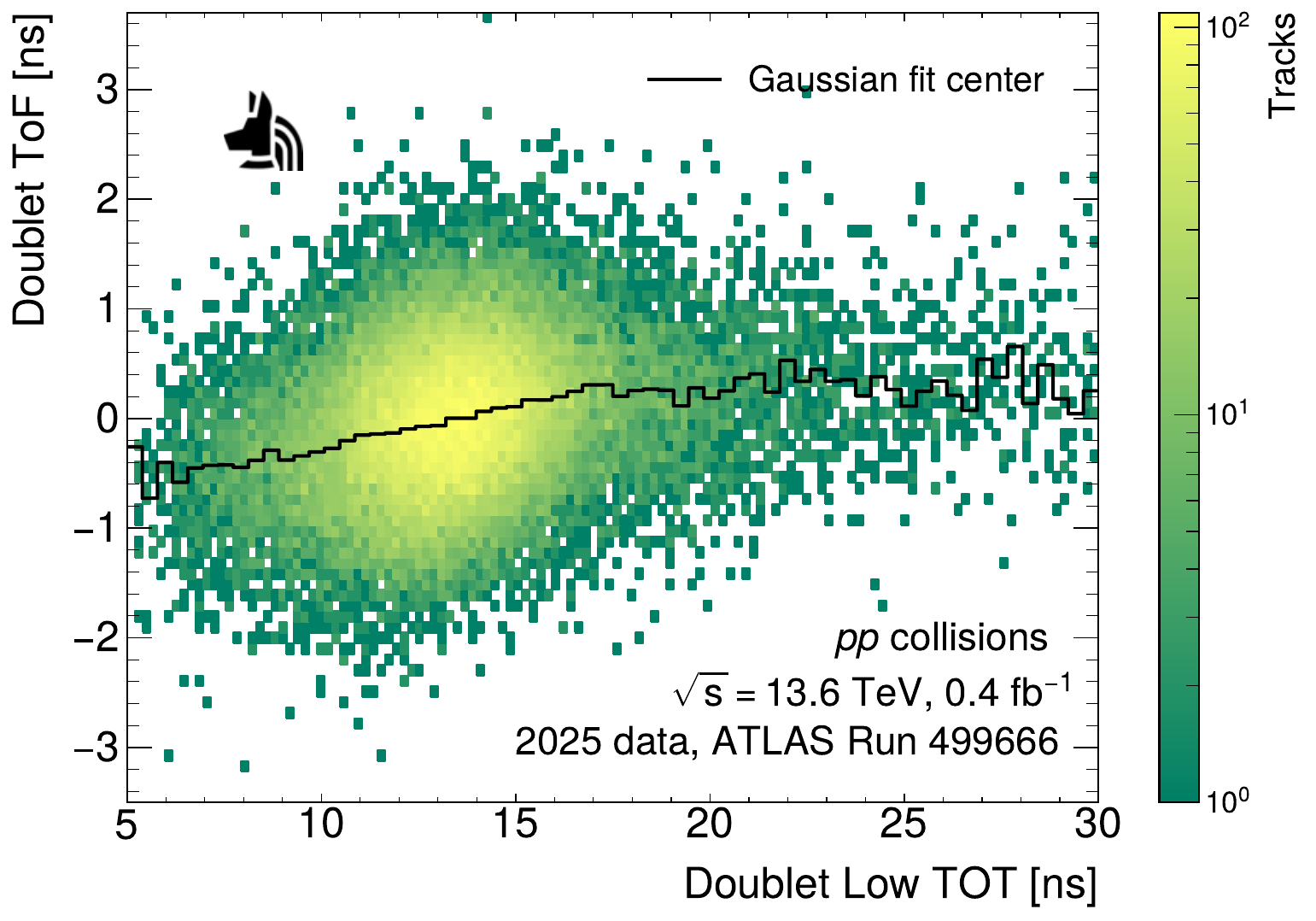}
\caption{
The distribution of the number of tracks as a function of the ToF between the two Doublet planes, and the Doublet Low TOT.
Also shown is the centre of the fit at each TOT (black solid line).
}
\label{fig:stepOneTOT}
\end{figure}

\begin{figure}[ht]
\centering
\includegraphics[width=0.45\textwidth]{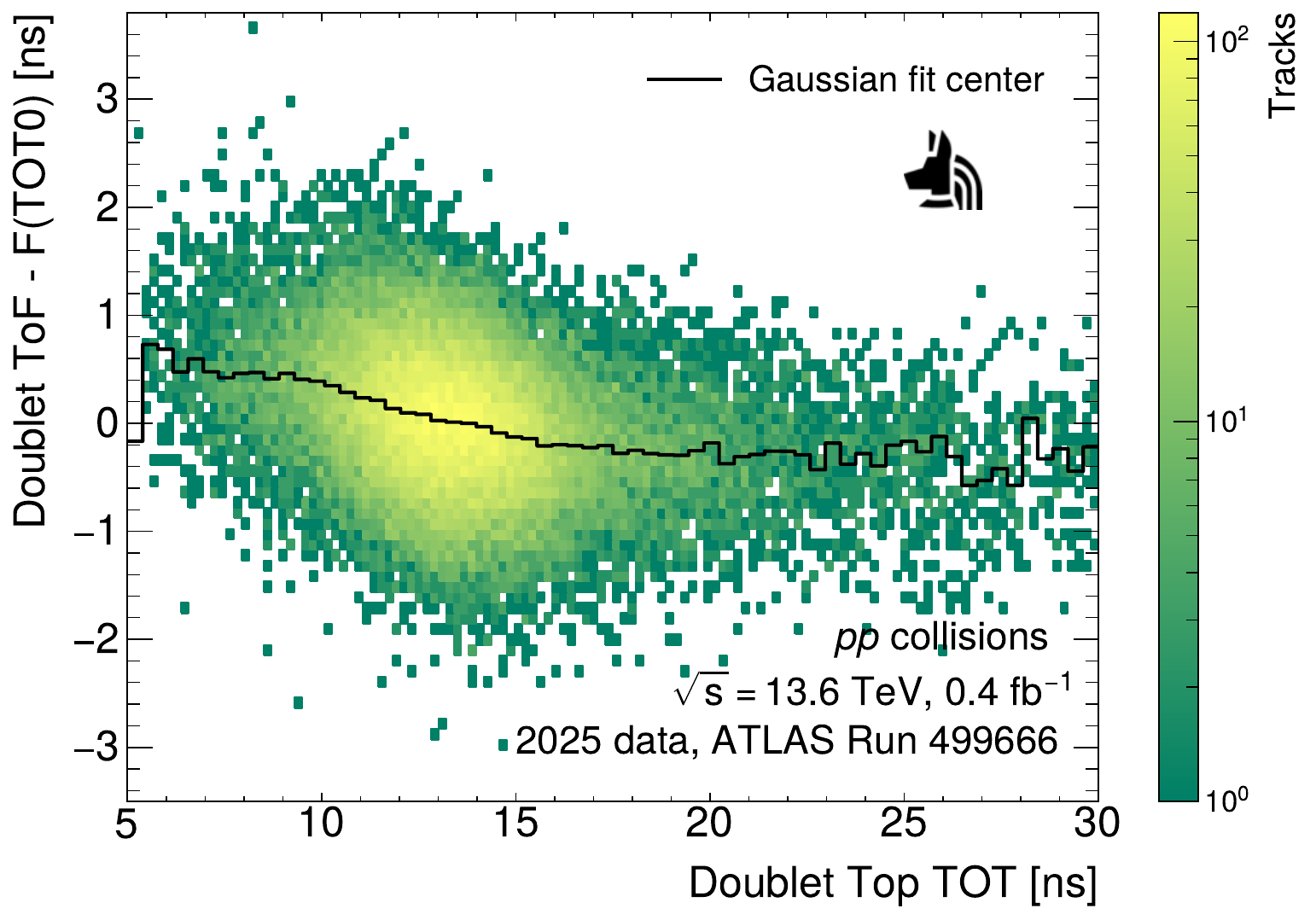}
\caption{
The distribution of the number of tracks as a function of the ToF between the two Doublet planes after correcting for F(TOT$_0$), and the Doublet Top TOT. 
Here, F(TOT$_0$) represents the calibration applied using the lower Doublet TOT.
Also shown is the centre of the fit at each TOT (black solid line). 
\\
} 
\label{fig:stepTwoTOT}
\end{figure}

\begin{figure}[ht]
\centering
\includegraphics[width=0.45\textwidth]{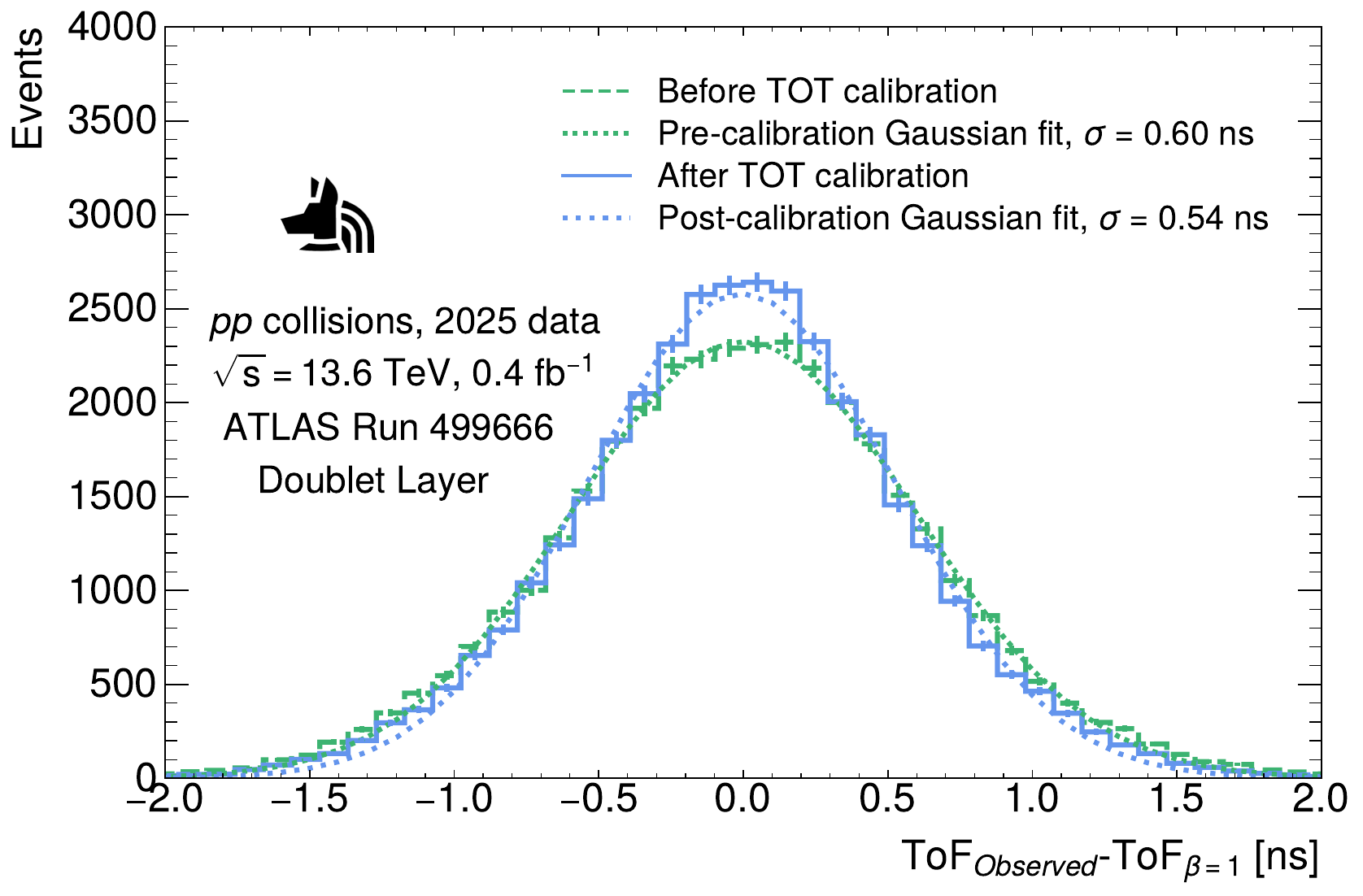}
\caption{The ToF difference relative to expectation in the Doublet RPC layer before and after applying the TOT calibration.} 
\label{fig:totCalib}
\end{figure}

Figures~\ref{fig:stepOneTOT} and~\ref{fig:stepTwoTOT} show an example distribution of ToF and TOT for the Doublet RPC layer in the two stages of the time-walk correction, along with the fitted means that are applied as the correction for each TOT. 
Figure~\ref{fig:totCalib} illustrates the impact of the TOT calibration on the inter-plane timing performance for those planes, demonstrating an improvement in the plane-to-plane time resolution from 0.60~$\pm$~0.003~ns to 0.54~$\pm$~0.003~ns following the application of the correction.

\section{Time Resolution}
With the full set of calibrations applied, the \proanubis time resolution is determined by comparing the hit times observed in planes with adjacent indices to the expected time difference for particles with $\beta~=$~1 using data collected during LHC collisions, where the expected time is calculated by dividing the physical distance between the planes by the speed of light.
As each measurement is made as a difference of two planes, its uncertainty is proportional to the quadratic sum of the uncertainties on the hit times in those planes.
To extract the resolution of each individual plane, the width of the time difference is found for all three pair-wise combinations of the planes in the Triplet layer and the resulting widths are used to compute the intrinsic timing resolution of the individual planes. A similar process is performed using the Singlet layer combined with the two planes of the Doublet layer. When using the Singlet and Doublet, the path length between the layers significantly reduces the angular acceptance of the track selection. Due to the larger relative sample size, only the Triplet layer is used for studies of the per-plane RPC timing resolution as a function of detector conditions.

\begin{figure}[ht]
\centering
\includegraphics[width=0.45\textwidth]{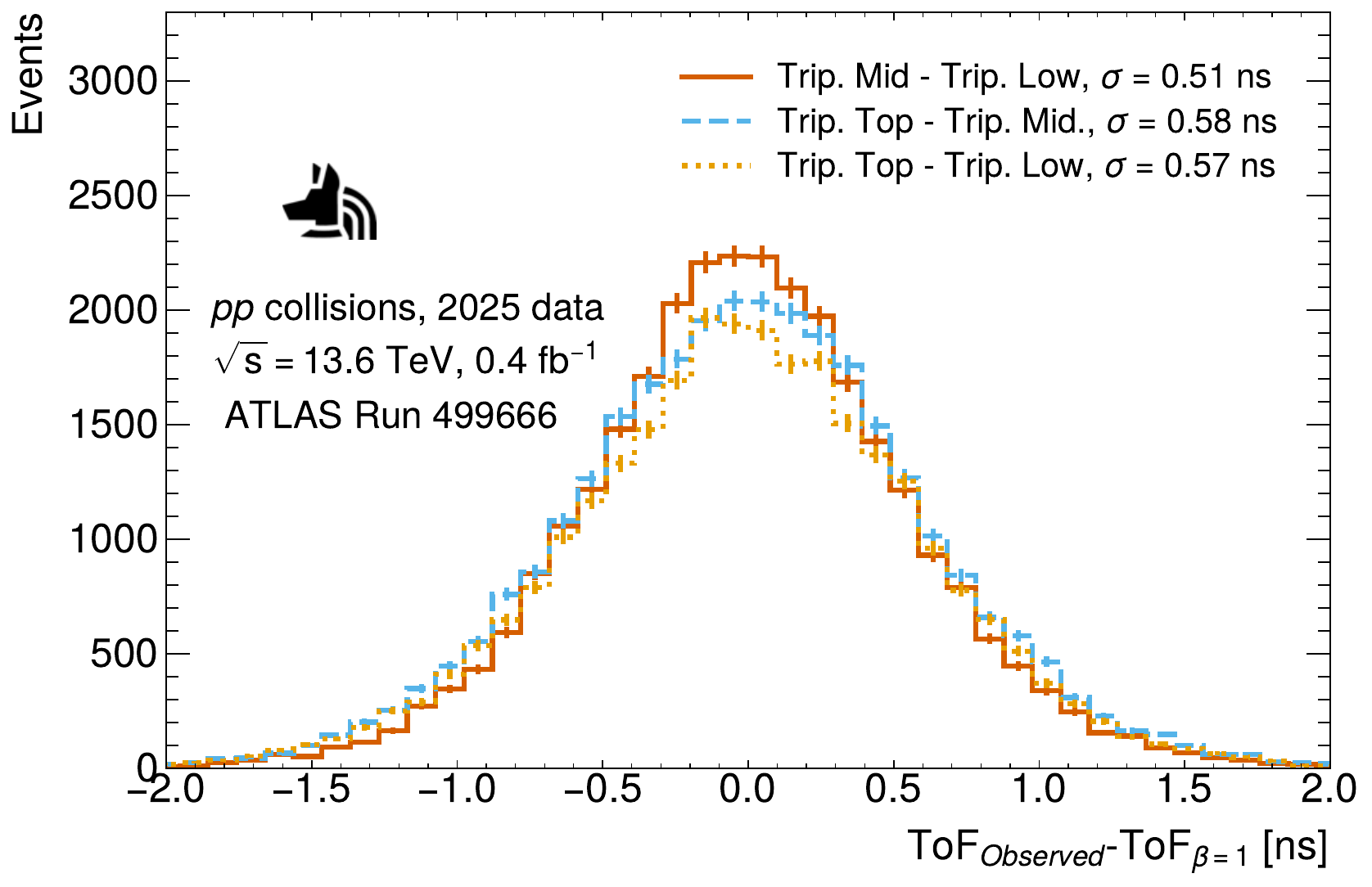}
\caption{Observed ToF relative to ToF for tracks with $\beta~$=~1 for all combinations of RPC planes in the Triplet layer.} 
\label{fig:tripletTimeRes}
\end{figure}

The ToF distributions for all combinations of RPC planes in the Triplet layer are shown in Figure~\ref{fig:tripletTimeRes}. 
From the three measured widths, individual plane time resolutions of 0.35~$\pm$~0.01~ns, 0.37~$\pm$~0.01~ns, and 0.45~$\pm$~0.01~ns are obtained for the lower, middle, and top RPCs of the Triplet layer, respectively. 
During data-taking the Triplet Top plane drew significantly higher current than the other two planes and was therefore operated at a reduced HV of 5.4~kV instead of the 5.6~kV applied to the other planes, which likely accounts for its correspondingly degraded time resolution. 

\begin{figure}[ht]
\centering
\includegraphics[width=0.45\textwidth]{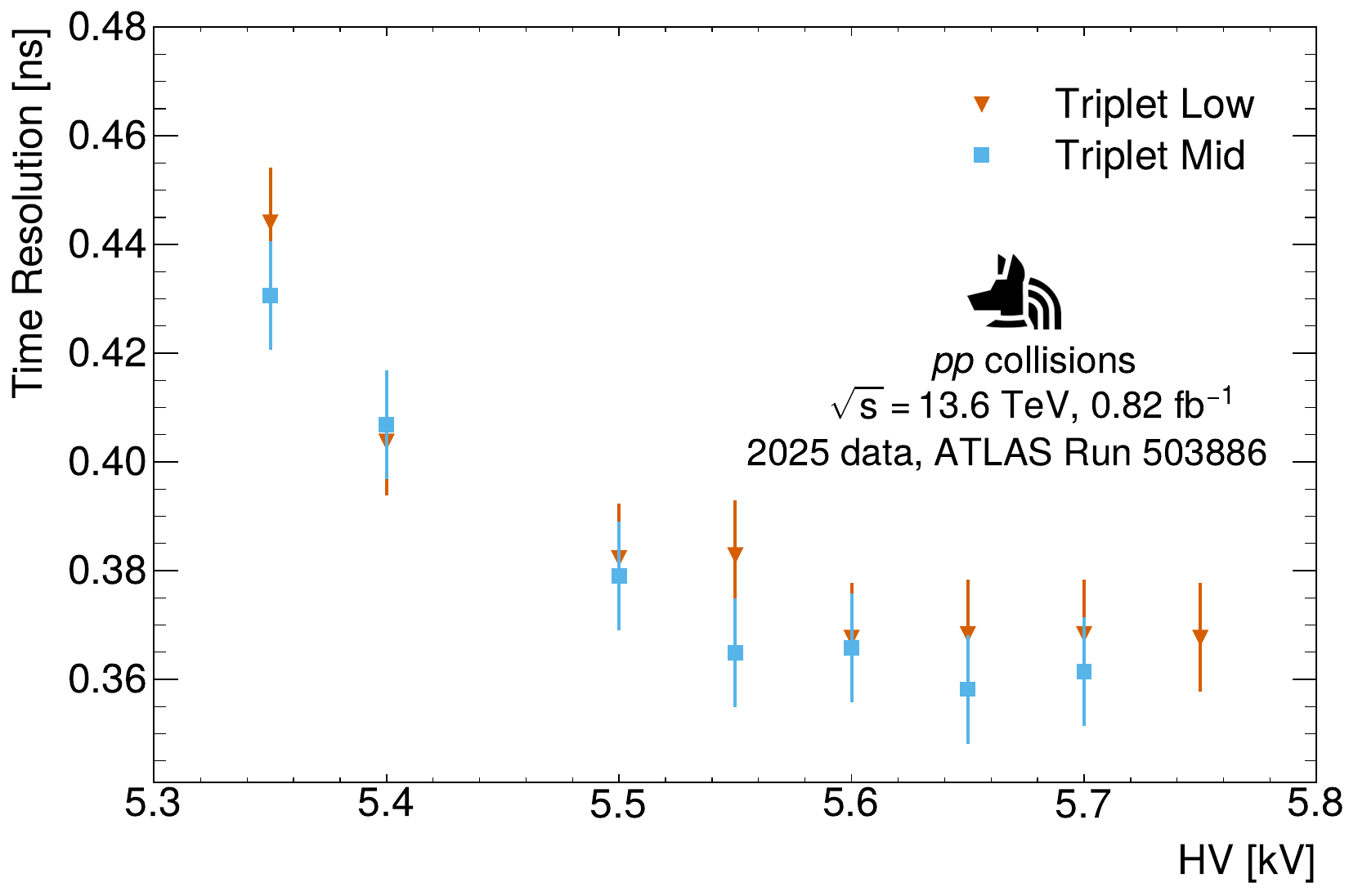}
\caption{Measured time resolution for the Triplet Mid and Low planes as a function of the high voltage applied.} 
\label{fig:timeResHVScan}
\end{figure}

To study the dependence of the RPC time resolution on the applied HV, the time resolution was measured as a function of input voltage, as shown in Figure~\ref{fig:timeResHVScan} for the Triplet Low and Mid planes.
The applied HV to the Triplet Top plane was not varied due to its higher current limiting the maximum safe HV.
For each measurement, the HV of a single plane was varied, while all other planes were maintained at their nominal operating voltages.
As expected, the time resolution improves as a function of HV up to about 5.6~kV, and then remains approximately constant at about 350~ps.
This would result in a per-triplet timing resolution of 200~ps for tracks with hits in each plane. 
Tracks with $\beta=~$1 traverse \proanubis in at least 3.6~ns, such that tracks with hits over two triplet layers containing at least two hits in each layer will have per-layer resolutions better than 250~ps and could reject downwards-going tracks with $>$7$\sigma$ accuracy.

A finer-grained analysis where each RPC is subdivided into regions in $\eta$ and $\phi$ has found that the majority of these regions achieve time resolutions close to 280~ps. 
The degraded resolution when using the entire RPC is caused by variations in the mean hit time that are correlated in $\eta$ and $\phi$ and thus are not accounted for by either the strip-wise timing calibrations or linear signal propagation corrections. 
Position-dependent variations in ToA of up to 1~ns are observed, potentially due to systematic effects in the individual calibrations as a function of $\eta$ and $\phi$ combining to produce 2D features, small non-uniformities in the surface of the RPC gas gaps, or correlations in the track angle and position in the detector biasing the timing performance. 
These variations are seen to be stable over month timescales, indicating that the time resolution could potentially be improved through more sophisticated calibration. 

\section{Summary and outlook} \label{Sec:Summary}
\vspace{-0.2cm}
The \proanubis detector, serving as a prototype for the proposed ANUBIS experiment, has been successfully installed in the ATLAS cavern and has collected data over the course of 2024 (104~\invfb) and 2025 (73~\invfb). Dedicated reconstruction algorithms have been developed to identify hits, clusters, and tracks in the recorded data, enabling an experimental evaluation of RPC detector performance in a realistic location similar to that of ANUBIS inside the ATLAS cavern. 

The \proanubis RPCs are found to have per-triplet-layer efficiency above 99$\%$ and per-RPC time resolutions below 500~ps. 
These results meet the target performance requirements of the ANUBIS detector design and demonstrate the suitability of RPC technology for large-area tracking and timing applications in long-lived particle searches.

Exploiting the time-synchronisation capabilities between \proanubis and the ATLAS experiment enables a range of future studies that combine information from both detectors. 
These include measurements of the precision of muon extrapolation between ATLAS and \proanubis, as well as particle time-of-flight measurements at the percent level.
Additionally, the ATLAS data can be selected for events with tracks passing through \proanubis in order to measure the reconstruction efficiency and resolution of the detector.
Lastly, the vertex reconstruction capability of \proanubis can be used in conjunction with ATLAS collision data to study the rate of punch-through vertices as a function of  hadronic activity that is closeby in $\eta,\phi$ space. 
Such measurements will provide an important validation of the background modelling assumptions used in the full ANUBIS proposal and its associated sensitivity projections.

\section*{Acknowledgements}
\vspace{-0.2mm}
We thank CERN for the very successful operation of the LHC and its injectors, the support staff at CERN, as well as our colleagues from the ATLAS Collaboration,  without whom proANUBIS could not be operated efficiently. 
We gratefully acknowledge the support of UKRI under the Future Leaders Fellowship scheme.

\printbibliography[title=References,heading=bibintoc]

\end{document}